# Unified chemical theory of structure and bonding in elemental metals

Yuanhui Sun[1,†], Lei Zhao[2,3,†], Chris J. Pickard[4,5], Russell J. Hemley[6], Yonghao Zheng[3], and Maosheng Miao[1,7]*

[1]*Department of Chemistry and Biochemistry, California State University Northridge, Northridge, California 91330, United States*

[2]*College of Chemistry and Chemical Engineering, Southwest Petroleum University, Chengdu, 610500, P.R. China*

[3]*School of Optoelectronic Science and Engineering, University of Electronic Science and Technology of China (UESTC), Chengdu 610054, P. R. China*

[4]*Department of Materials Science & Metallurgy, University of Cambridge, 27 Charles Babbage Road, Cambridge CB3 0FS, United Kingdom*

[5]*Advanced Institute for Materials Research, Tohoku University 2-1-1 Katahira, Aoba, Sendai, 980-8577, Japan*

[6]*Departments of Physics, Chemistry, and Earth and Environmental Sciences, University of Illinois Chicago, Chicago, IL 60607, United States*

[7]*Department of Earth Science, University of California Santa Barbara, California 93106, United States*

†These authors contributed equally to this work.
*Corresponding author. Email: mmiao@csun.edu

**ABSTRACT**

Most elemental metals under ambient conditions adopt simple structures such as BCC, FCC and HCP in specific groupings across the Periodic Table, and on compression many of these elements undergo transitions to surprisingly complex structures, including "open" and low-symmetry phases not expected from conventional free-electron based theories of metals. First-principles calculations have been able to reproduce many observed structures and transitions, but a unified, predictive theory of bonding that underlies this behavior is not yet in hand. We propose a remarkably simple theory based on large-scale high-throughput calculations of the elements over a broad range of thermodynamic conditions. The results broaden the conventional concept of metallic bonding with a new perspective that both explains the stability of different simple structures as well as lower symmetry phases arising from electron localization at interstitial sites in these structures. The success of this simple framework reveals the important role of chemical interactions in governing the structures and electronic properties of simple metals.

**INTRODUCTION**

The modern theory of metals may be considered one of the great successes of quantum mechanics applied to solids [1,2]. This development over many decades has given rise to the modern band picture in physics, as implemented in numerous highly successful band-structure (e.g., density-functional theory) methods that are now able to reproduce experimentally measured ground properties of alkali, alkaline earth, and transition metals. Despite this success, our understanding of the structural variations of elemental metals across the Periodic Table and as a function of thermodynamic conditions remains unsatisfied [2–5] within this "physics" (band) picture. In parallel with this dominant view of the filed, alternative perspectives that begin with and focus on local bonding considerations were explored [6,7]. This "chemical" (bond) approach has been revisited periodically in various contexts in later years [8–11], for example in the treatment of various types of electron localization. Here, we show that the electron occupation of the interstitial local orbitals and their corresponding chemical interactions are the key factors that determine the structures of elemental metals and their evolutions under pressure.

Elemental metals are among the simplest solid forms of matter. However, their structures show an intricate variation across the periodic table and many of them undergo transitions from high symmetry to complex structures on compression [2–5]. A full theory of metals requires a unified framework to understand and predict all their structures, transitions, and stabilities [4,5]. Under ambient conditions all alkali metals, Ba, and group 5 and 6 transition metals adopt a BCC structure; Be, Mg, group 3, 4, 7, and 8 transition metals (except Mn, Fe), Zn, and Cd, adopt an HCP structure; Ca, Sr and most late transition metals (except Co) adopt an FCC structure (Fig. S1). Furthermore, on compression, alkali metals transform into an FCC structure, and Ba into an HCP structure, whereas many alkaline earth metals and group 4 transition metals including Mg, Ca, Sr, Zr, and Hf into a BCC structure. While many transition metal structures remain "simple" up to very high pressures, the so-called "simple" *s*-block metals may pass through a series of complex structures with large interstitial sites and low symmetry [5].

Modern band-structure methods can reproduce and have even predicted the existence and stability of many of the structures [12–21], but the origin of the phenomena and underlying mechanisms are not understood. Conventional band-structure approaches emphasize the delocalization aspect of the electrons in metals [2,22] and account for physical features such as band filling [14], Fermi surface and Brillouin zone topology [16–19], *s-d* transfer [23–25], etc. These approaches approximate various features of the electronic structure, and often are limited to a group of metals or phenomena. For example, the concept of Fermi surface nesting [18,19,26] has been used to explain structural changes in Li and K under pressure, but not in Na, because the Fermi surface of the latter remains spherical [21]. More recently, the existence of dynamical instabilities in which lattice becomes unstable with respect to atomic displacement has been investigated to explain structural changes in metals [19,21,23]. However, instead of revealing the mechanism, these results actually added a new question to the conundrum, *i.e.* why is the thermodynamic instability of simple metal structures often accompanied by a dynamic instability? As a result, a predictive framework for structural patterns and their evolution remains to be developed. Here we show that the ambient and high-pressure structures of most metals can be explained surprisingly well by a simple theory, if we change from the "physics" band-structure point of view to a "chemical" perspective focusing on the electron occupations of quasi-atom orbitals and their bonding interactions.

**RESULTS AND DISCUSSION**

**Electron Localization and Close-Packed Structures**

To understand and predict the full range of structures adopted by simple metals over a wide range of conditions, we first need to depart from the view of electrons occupying nearly-free delocalized states towards a picture of occupying local orbitals centered not at the atoms but at interstitials. The occupation of these interstitial orbitals, or quasi-atom orbitals, can be used to explain and predict the formation of high-pressure electrides (HPE) [11]. In HPE, such as *hP*4 Na at 200 GPa (Figs. 1 Aa – Ad), the electron localization functions (ELFs) and the electron density show distinct maxima in the interstitial regions resulting from to the occupation of quasi-atom orbitals, and can therefore be viewed as anionic species in a solid compound [27–29].

This quasi-atom scenario can be extended to ambient or low-pressure conditions in cases where the interstitial orbitals are partially occupied, and the metals are viewed as "electron" compounds, *i.e.* compounds consisting of quasi-atoms. For example, the ELFs of BCC Na at ambient pressure show maxima at the tetrahedral sites ($E^T$) (Fig. 1Ba), and FCC Na at 70 GPa show maxima at the octahedral ($E^O$) and $E^T$ sites (Fig. 1Ca). Accounting for the quasi-atoms, metals in the FCC and HCP structures resemble binary compounds in NaCl and anti-NiAs structures, and BCC is isostructural to compounds such as sodalite $CaH_6$ [30] and $SrB_2C_4$ [31]. Both BCC and FCC Na show very weak density maxima at interstitial sites and are not electrides (Fig. 1, Bb and Cb). On the other hand, the ELF values and patterns are consonant with the electron density differences between metal lattices and free atoms (Fig. 1, Bc and Cc), and are in accord with the occupation of quasi-atom orbitals (Fig. 1, Bd and Cd). Electron localization in interstitial regions has been noted in several cases of metals and metal clusters [9,24,32–36], but the occupation of the local quasi-atom orbitals and the significance of the corresponding chemical interactions have not been explored. We will show that they lead to a universal mechanism that explains the structural trends of metals.

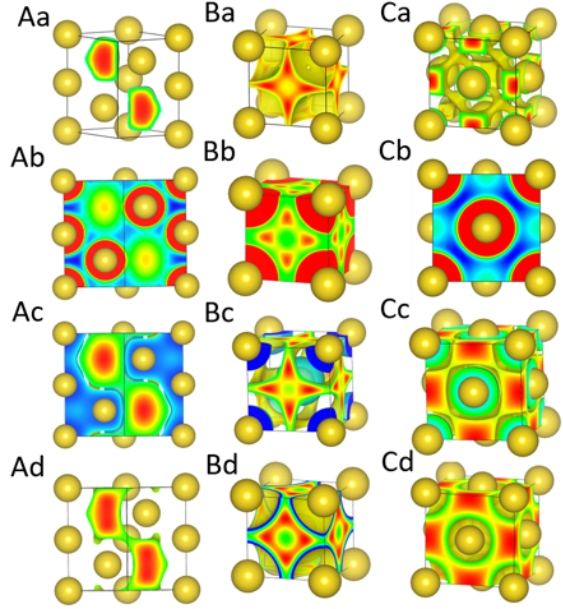

**FIG. 1. Electron localization and subset interactions in metals. (Aa – Ad)** ELF, charge density, the charge density difference between metal lattice and metal atoms, and a crystal orbital at the G point of Na in *hP*4 structure at 200 GPa. **(Ba – Bd)** The same as **(Aa – Ad)** for Na in BCC structure at ambient pressure. **(Ca – Cd)** The same as **(Aa – Ad)** for Na in FCC structure at 70 GPa. The color code in **(Ab, Bb, Cb)** is adjusted to signify the maximum values.

Interstitial electron localization occurs in various metal lattices, despite whether they are the stable structures of the metals or not (Fig. S2). Their locations and strengths, as revealed by ELF, show a strong dependence and orderly evolution across various elements, lattices, and volumes (Figs. S3 to S9). With decreasing lattice constants electron localizations tend to change from sites with fewer neighboring atoms to sites with more neighboring atoms. For example, in a simple cubic (SC) lattice, the electron localization sites tend to change from the edge centers with 2 neighboring atoms to face centers with 4 neighboring atoms, and then to body centers with 8 neighboring atoms (Fig. S3A). Similarly, electron localization in an FCC structure tends to change from bond centers with 2 neighboring atoms, to tetrahedral sites with 4, and then to octahedral sites with 6 neighboring atoms (Fig. S3C).

The electron occupation of the localized quasi-atom orbitals and the corresponding chemical interactions between these orbitals determine the structure preference between FCC and HCP of a metal. From the standpoint of quasi-atoms, the major difference between the two structures is that the tetrahedral interstitials form pairs in HCP (Fig. S10A) but form a lattice in FCC (Fig. S10B). Electrons in close-packed Be and Mg highly localize at the $E^T$ sites (Figs. 2, Aa and Ab), which causes very strong chemical $E^T$-$E^T$ pair interactions in HCP that stabilize it (Fig. 2B). Similar pair interactions of quasi-atoms have been shown before in very different circumstances. [37,38] Consequently, *c/a* ratios in Be and Mg are 1.568 and 1.626, both smaller than the ideal value of 1.633. In contrast, electrons in close-packed Ca (Fig. 2Ac) and Sr (Fig. 2Ad) localize mainly at the $E^O$ sites (Fig. 2C) due to their weaker ion-electron interactions. Comparing with ionic compounds and counting the localized electrons at $E^O$ sites as partially charged anions, FCC corresponds to the NaCl structure whereas HCP corresponds to the anti-NiAs structure. The electrostatic energy is lower in FCC because the Madelung constant in the NaCl structure is 1.748, which is significantly larger than that of 1.693 in anti-NiAs structure, which explains why Ca and Sr prefer FCC.

Like Be and Mg, electrons in close-packed transition metals of Groups 3 – 8 tend to localize on the $E^T$ sites, causing their HCP structures stabilized by the strong pair interactions (Fig. 2D). Accordingly, the $c/a$ ratios of their HCP structures are also below 1.633. Compared to alkaline earth metals, the octahedral site localizations are also large in the early transition metals, but the effect is not significant enough to reverse the FCC-HCP stability order. For the late transition metals, the ELF values become insignificant, indicating a weaker $E^T$-$E^T$ pair interaction effect, and the FCC structure is stable for these metals (Fig. 2D).

Zn and Cd appear as a radical departure from the general trend, as they adopt the HCP structure with $c/a$ ratios of 1.856 and 1.885, significantly higher (rather than lower) than the ideal value. This unusual behavior is due to the unique electron localization in these two elements. Compared with Be and Mg, electrons in HCP Zn and Cd highly localize not only on the $E^T$ sites but also on the triangular sites in the hexagonal plane (Fig. 2E). These electrons show also strong bonding with the neighboring $E^T$ sites. On the other hand, the chemical interactions between the localized electrons in neighboring hexagonal layers are much weaker. Thus, HCP Zn and Cd behave like layered compounds and show exceedingly large $c/a$ ratios.

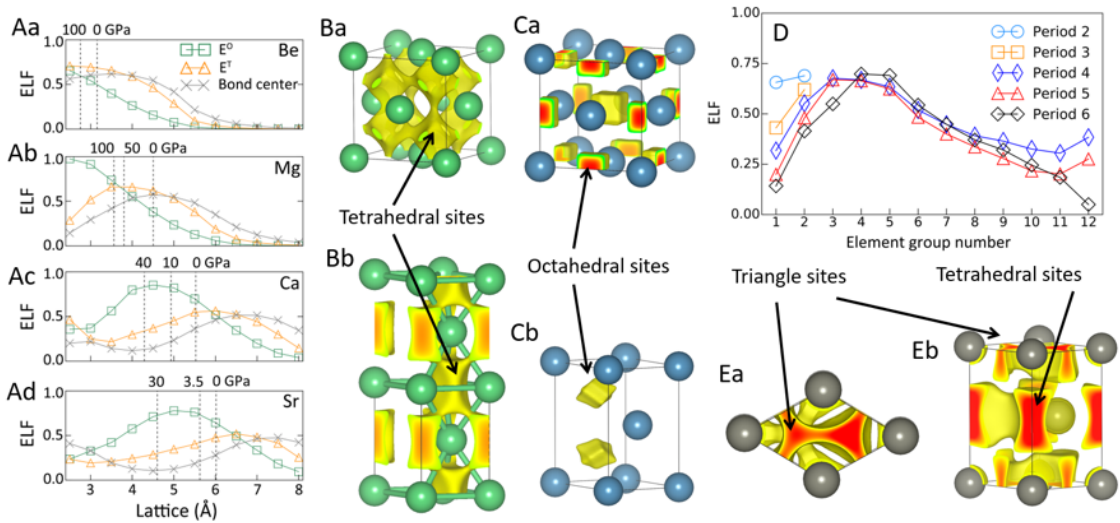

**FIG. 2. ELF distributions in metals.** (**Aa – Ad**) ELF graphs showing ELF values at the high symmetry points in FCC Be, Mg, Ca, and Sr. The vertical lines show the pressures of the FCC lattices. (**Ba – Bb**) ELFs of Be in FCC and HCP structures. They show large electron localization at $E^T$ sites that form an SC lattice in FCC and pairs in HCP structures. (**Ca – Cb**) ELFs of Ca and Sr in FCC and HCP structures. They show large electron localization at $E^O$ sites. (**D**) The ELF values at the $E^T$ sites of various metals in FCC structure. (**Ea – Eb**) The top and the side views of ELF of Zn in HCP structure.

**Subset Interactions and Stability of BCC**

The stability of BCC is governed by a more intricate mechanism, but as we show below, is still based on the principles discussed above. To reveal that, we need to split the metal lattice into two equivalent sublattices (subsets) (Fig. S10), e.g., BCC to two SC lattices (Fig. S10C). While two SC lattices interpenetrate to form a BCC, the high symmetry points of one SC lattice including the lattice point (L), the edge center (E), the face center (F), and the body center (B) become the B, F, E, and the L points of another SC, respectively. Notably, the $E^T$ point of a BCC lattice corresponds to geometrically identical quarter-center points (denoted as $E^T_{BCC}$) in both SC sublattices with coordinates of (0.25, 0.5, 0) (Fig. S10C).

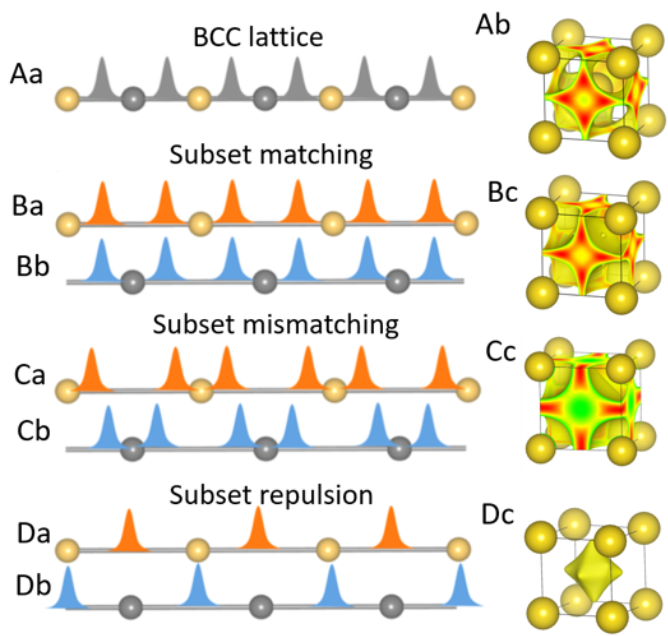

**FIG. 3. Subset interactions in metals.** (**Aa**) Schematic diagram of electron localization in metal lattices. (**Ab**) ELF shows the electron localization in BCC Na at 5 GPa. (**B**) Schematic of subset matching in a metal lattice. (**Ba – Bb**) Two sublattices that perfectly match that of the central metal lattice as shown in (**Aa**). (**Bc**) SC Na as a sublattice of BCC Na at 5 GPa. (**C**) Schematic of partial subset matching. (**Ca – Cb**) The electron localization in two sublattices that partially match that of the metal lattice. (**Cc**) SC Na as a subset of BCC Na at 0 GPa. (**D**) Schematic of subset repulsion. (**Da – Db**) Two sublattices that repulse each other. (**Dc**) SC Na as a sublattice of BCC Na at 70 GPa.

The subsets of a metal lattice might impose strong interactions to each other due to the match or mismatch of their interstitial electron localizations (Fig. 3). These interactions can be summarized into three representative cases, including a matching, a mismatching, and a repulsing case. In a perfect matching case (Fig. 3B), the interstitial electron localization in each sublattice reproduces that of the whole, therefore ELF patterns and WFs of all lattices (the sublattices and the whole) match. As a results, the chemical interactions between the sublattices enhance the stability of the structure. In a mismatching case (Fig. 3C), the electron localization of each sublattice either shifts away or contributes only part (half) of the electron distribution of the whole lattice, usually corresponding to a less stable structure. In a repulsing case (Fig. 3D), the electrons in one sublattice localizes on the atom sites of another sublattice (counter-atom), imposing repulsive forces due to interactions with the ion core, and destabilizing the corresponding metal structure. Compared with FCC and HCP, BCC can be divided into two sublattices that do not contain nearest neighbors (Fig. S10C), in which case subset interaction effects are the strongest and the metal will adopt BCC if its lattice constant is at or very close to the matching point.

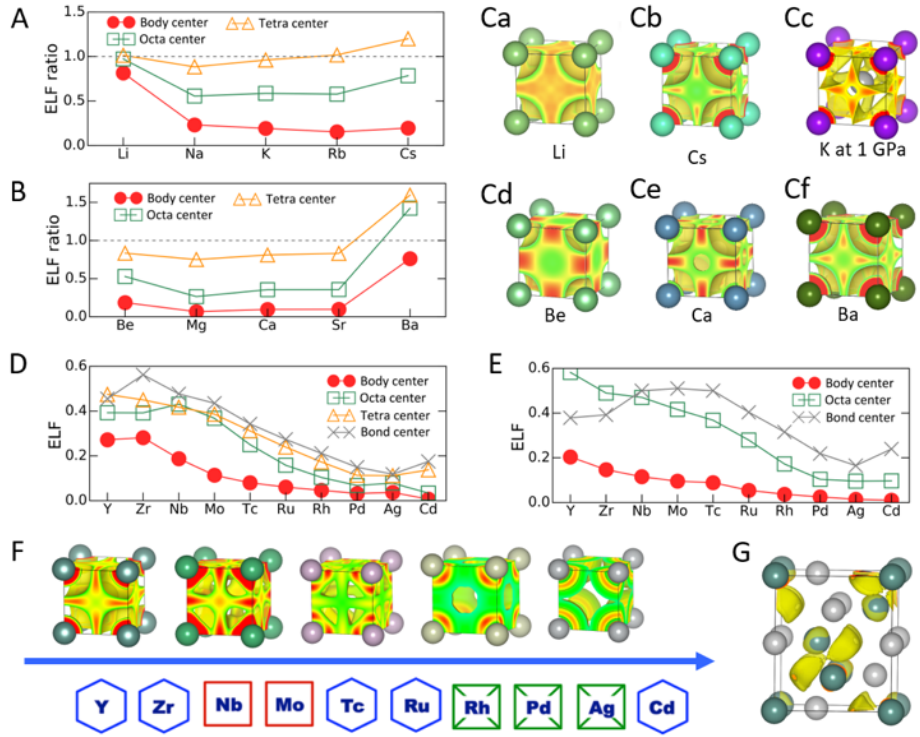

**Fig. 4. Stability of BCC structure at zero pressure unless otherwise specified.** (**A**) ELF/ELF(edge center) ratios in the sublattices of alkali metals having bcc structures. (**B**) ELF/ELF(edge center) ratios in the sublattices of alkaline earth metals in BCC structure. (**C**) Sublattice ELF of selected alkali and alkaline metals having BCC structures, including Li, Cs, K at 1 GPa, Be, Ca, and Ba. (**D**) Sublattice ELF values at high symmetry points of 5$^{th}$-row transition metals in the BCC structure. (**E**) Sublattice ELF values at high symmetry points of 5$^{th}$-row transition metals in FCC. (**F**) ELF of 4$d$ transition metals BCC sublattice (SC). From left to right are Y, Nb, Tc, Rh, and Ag. (**G**) ELF of Y HCP sublattice. The green and the grey balls show the atoms in two sublattices. The ELF of the sublattice has maxima at the E$^O$ sites of the HCP, showing sublattice matching.

The ELFs of alkali metals in BCC structures show either maxima (in the case of Li and Cs) or large values at the $E^T_{BCC}$ sites (in the case of Na, K, and Rb) (Figs. 4A, 4Ca – 4Cc). In the latter case, a small compression causes a shift of ELF maxima from edge centers to $E^T_{BCC}$ sites. For example, Na, K, and Rb reach perfect sublattice matching at 5.0, 1.0, and 0.7 GPa (Fig. 4Cc). In contrast, the BCC structures of most alkaline earth metals are well away from the sublattice matching point, and their ELF at $E^T_{BCC}$ sites are distinctly smaller than those at the edge centers (Figs. 4B, 4Cd, 4Ce). Therefore, most alkali metals adopt the BCC structure, whereas most alkaline earth metals adopt close-packed structures at or near ambient conditions. Furthermore, Li and Ba have distinctly different electron localization features compared with other elements in the same group. In contrast to other alkaline earth metals, the sublattice ELF of BCC Ba maximizes at $E_{BCC}^T$ sites (Fig. 4Cf), consistent with its stability in the BCC structure. On the other hand, although Li BCC is close to the sublattice matching point, its ELF at the body centers is very high, which gives rise to strong sublattice repulsions (Fig. 4A). Therefore, lattice matching and repulsive effects coexist in BCC Li at zero pressure, compromising its stability.

For transition metals, the electron localization in interstitial regions as quantified by the ELF decreases with increasing number of $d$ electrons because of the increase in the nuclear attraction potential (Figs. 4D – 4F, Fig. S11). For early transition metals such as Y and Zr, ELF values are high at all high symmetry points, including the body centers, indicating strong sublattice repulsions that destabilize the BCC structure (Figs. 4D, 4F, S11B, S11D). On the other hand, the sublattice of their HCP structures exhibit ELF maxima locating close to the octahedral sites of the original HCP structure, indicating strong sublattice matching that stabilizes HCP (Figs. 4E, 4G, S11F). For elements in Groups 5 and 6 (*e.g.,* Nb and Mo), the BCC sublattice ELF at the body centers decreases more significantly than other points, greatly reducing sublattice repulsions (Figs. S11B, S11E). Furthermore, their BCC lattices are close to the sublattice matching point, gaining notable stability against other structures. Later transition metals exhibit lower interstitial localization and therefore weaker sublattice interactions and tend to adopt close-packed structures. On compression, some transition metals such as Zr and Hf, transform to BCC [5] because the increasing localizations at sublattice face centers move it closer to sublattice matching point (Figs. S12A, S12B). Group 5 and 6 transition metals remain in the BCC structure up to very high pressure [39] as a result of very low compressibility and weak dependence of the ELF on lattice constant (Figs. S12C, S12D).

**Structures of Metals under Pressure**

Any theory of metals must also explain their structure changes under pressure. It has long been known, for example, that the alkali metals transform from BCC to FCC, whereas alkaline metals change from HCP or FCC to BCC. Sodium crystallizes in the BCC structure at ambient pressure and transforms to FCC at 65 GPa, and to *cI*16 at 104 GPa [5,27,40]. The perfect matching point of BCC Na occurs at 5 GPa, at which point the ELF maximizes at the $E_{BCC}^T$ of both SC sublattices (Figs. 5, A and C). With increasing pressure, the electron localization in the SC lattice shifts from $E_{BCC}^T$ to F and then to C sites, lowering the matching effect and enhancing the sublattice repulsions. At higher pressure of 65 GPa, the sublattice repulsion is strong enough to destabilize BCC. On the other hand, the electrons in one sublattice of FCC Na localize mainly between two Na atoms of the other sublattice, consistent with the stability of FCC Na at this pressure (Figs. 5, B and C). At a higher pressure of 110 GPa, FCC Na develops strong enough subset repulsions to destabilize it (Fig. 5C). Similar trends are also found for other alkali metals (*e.g.,* K; Fig. S13).

Sublattice interactions also explain the fact that most alkaline earth metals transform from FCC or HCP to BCC under pressure, a trend that is opposite to that of alkali metals. Alkaline earth metals, except Ba (Fig. S14), exhibit large ELF values at the edge centers of SC lattice (BCC sublattice) at ambient and low pressures (*e.g.,* Ca; Fig. S15) [5,41]. The BCC structure has no advantage over FCC in subset matching. On compression, however, the ELF maxima change from edge centers to the face centers in a specific pressure range (*e.g.,* Ca at 10 GPa; Fig. S15), the ELF maximum locates at $E_{BCC}^T$, showing perfect subset matching, which drives the phase transition from FCC to BCC. At around 30 GPa, BCC loses its stability due to the sublattice repulsion (Fig. S15). In contrast to other alkaline earth metals, Ba shows a reverse structure transformation under pressure from BCC to HCP. This can be understood by the observation of a matching point for its BCC structure at zero pressure which is removed with increasing subset repulsion on compression (Fig. S14).

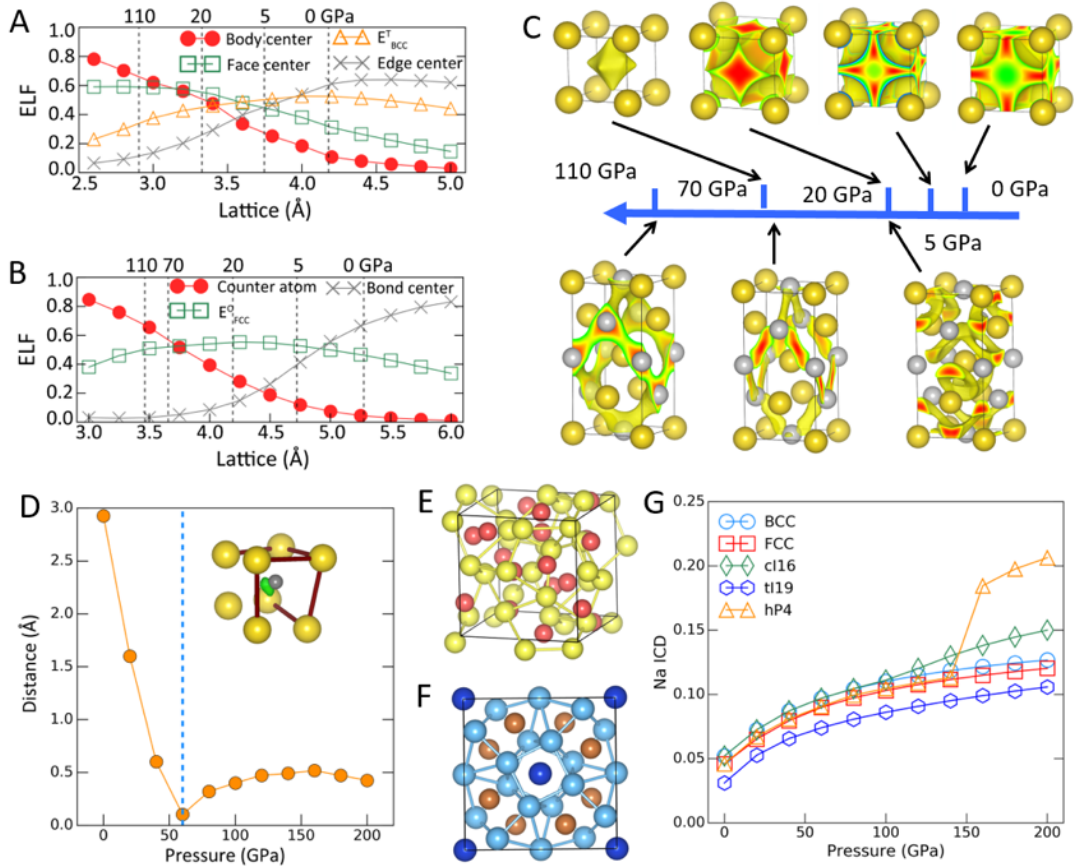

**FIG. 5. Metals under pressure.** (**A**) The ELF values at the high symmetry points of SC Na with lattice parameters of 2.5 to 4.5 Å. The vertical lines show the pressures of BCC with the same lattice parameters. SC is the sublattice of BCC. (**B**) The ELF values at the high symmetry points of a Na lattice that is the sublattice of FCC, with lattice parameters from 3.0 to 6.0 Å. The vertical lines show the pressures of the corresponding FCC lattice. (**C**) Schematic evolution of ELF of BCC and FCC Na sublattice with increasing pressure. (**D**) Reduction of the sublattice repulsion in $cI$19 Na due to its deviation from an ideal BCC structure. The inset shows the sublattice lattice and its ELF maximum at the body center. The plot shows the distances between the ELF maximum (green spot) generated by one sublattice (yellow balls) and the closest Na atom from the other sublattice (grey ball). At about 60 GPa, $cI$19 reduces to a perfect BCC structure, and the ELF maximum starts to shift away from body centers to face centers and then to edge centers with decreasing pressure. (**E**) $Ba_4As_3$ in $I\bar{4}3d$ structure, in their yellow and red balls represent the Ba and the As atoms. (**F**) Structure of $Ti_5CuSb_2$, where the light blue balls represent Ti atoms, and the dark blue and brown balls represent Cu and Sb atoms, respectively. (**G**) The calculated integrated charge differences of Na atoms in various structures as functions of pressure (see Methodology Section).

Sublattice repulsions also cause dynamical instabilities in high symmetry structures and the formation of the "open" structures of alkali and alkaline earth metals at high pressure: by moving the atoms away from the high symmetry points, these structures can avoid subset repulsion and lower their energy. For example, the $cI$16 structure of Na that is stable from 104 GPa to 117 GPa can be viewed as distorted BCC. Splitting $cI$16 into sublattices analogues to the corresponding BCC, the ELF maxima of one sublattice are no longer located on the lattice points of the second sublattice. The distance between ELF maxima and the lattice point changes with pressure (Fig. 5D). For Na, at 110 GPa, this distance is about 0.5 Å, large enough to effectively avoid sublattice repulsion.

Electron localization at the interstitial sites is a key feature of "open" structures under pressure. While including the quasi-atoms whose orbitals are partially occupied, the complex high-pressure structures are also analogous to solid compounds. For example, $cI$16 Na can be viewed as a binary compound with $A_4B_3$ composition and $I\bar{4}3d$ structure, such as $Ba_4As_3$, in which As atoms occupy the positions of the quasi-atoms (Fig. 5E, S16A, S16B). Similarly, the $tI$19 Na host-guest structure is isostructural to the ternary compound $Ti_5CuSb_2$, in which Cu and Sb play the roles of two types of quasi-atoms inside $tI$19 structure (Fig. 5F, S16C, S16D). The actual charge transfer from metals to quasi-atoms in all phases is comparable, as calculated by integrated charge difference (ICD), especially at lower pressures (Fig. 5G, S16G). Above 120 GPa, charge transfer increases in high-pressure structures because of changes in atomic versus quasi-atomic orbitals. Heavier alkali metals, alkaline earth metals, and transition metals exhibit different degrees of pressure-induced charge transfer into interstitial sites as a result of competition with $d$ orbital charge transfer. The large charge transfer into both interstitial sites to $d$ orbitals also reduces subset repulsions, which causes the presence or reappearance of high symmetry structures such as HCP and $hP$4 at still higher compressions.

The framework described above also applies to $p$-block metals such as Al, which further reveals that it is not the density but the electron occupation of local orbitals that governs structural preference and its evolution under pressure. For example, FCC Al is a good metal with high electron density at ambient pressure with weak electron localization at its interstitial sites; the metal undergoes structural transformations as $s$-block metals but with a much larger pressure scale due to its weaker dependence of change of electron localizations with pressure (Fig. S17). The surprising generality of the framework presented above arises from the realization of the importance of electron localization and occupation of local quasi-atom orbitals in metals and the impact of chemical interactions in determining structural stability.

## ACKNOWLEDGMENTS

**Funding:** M.M. and Y.S. acknowledge the support of NSF under grant No. DMR 1848141, and computational resources provided by XSEDE (TG-DMR130005). M.M. also acknowledges the support of ACS PRF 50249-UNI6 and the support of California State University Research, Scholarship and Creative Activity (RSCA) award. R.J.H. acknowledges support from NSF (DMR-2104881) and DOE-NNSA (DE-NA0003975). **Author contributions:** M.M. proposed the mechanism and designed the research. Y.S. and L.Z. made equal contributions and conducted all the calculations. M.M and Y.S. analyzed the results, and other authors contributed via discussions. M.M. wrote the first draft and all authors wrote the manuscript together. M.M. and Y.S. plotted the figures. **Competing interests:** The authors declare no competing financial interest. **Data and

Supplementary Material for

# Unified chemical theory of structure and bonding in elemental metals

Yuanhui Sun[1,†], Lei Zhao[2,3,†], Chris J. Pickard[4,5], Russell J. Hemley[6],
Yonghao Zheng[3], and Maosheng Miao[1,7]*

[1]*Department of Chemistry and Biochemistry, California State University Northridge, Northridge, California 91330, United States*

[2]*College of Chemistry and Chemical Engineering, Southwest Petroleum University, Chengdu, 610500, P.R. China*

[3]*School of Optoelectronic Science and Engineering, University of Electronic Science and Technology of China (UESTC), Chengdu 610054, P. R. China*

[4]*Department of Materials Science & Metallurgy, University of Cambridge, 27 Charles Babbage Road, Cambridge CB3 0FS, United Kingdom*

[5]*Advanced Institute for Materials Research, Tohoku University 2-1-1 Katahira, Aoba, Sendai, 980-8577, Japan*

[6]*Departments of Physics, Chemistry, and Earth and Environmental Sciences, University of Illinois Chicago, Chicago, IL 60607, United States*

[7]*Department of Earth Science, University of California Santa Barbara, California 93106, United States*

†These authors contributed equally to this work.
*Corresponding author. Email: mmiao@csun.edu

## Materials and Methods

**Solid-state density functional calculations.** The underlying first-principles density functional theory (DFT) calculations were carried out by using the plane-wave pseudopotential method as implemented in Vienna *ab initio* Simulation Package (VASP) [42, 43]. The electron-ion interactions were described by the projector augmented wave pseudopotentials [44, 45] and the used valence electrons are listed in Table 1. We used the generalized gradient approximation formulated by Perdew, Burke, and Ernzerhof [46] as exchange-correlation functional. A kinetic energy cutoff of 520 eV was adopted for wave-function expansion. The *k*-point meshes with interval smaller than $2\pi \times 0.03$ Å$^{-1}$ for electronic Brillouin zone to ensure that all enthalpy calculations converged within 0.02 eV/atom. The high-throughput first-principles calculations were performed by using the Jilin Artificial-intelligence aided Materials-design Integrated Package (JAMIP), which is an open-source artificial-intelligence-aided data-driven infrastructure designed purposely for computational materials informatics [47].

**Table S1.** The valence configurations of the pseudopotentials used in our solid-state DFT calculations.

| Li | Be | | | | | | | | | |
|---|---|---|---|---|---|---|---|---|---|---|
| $1s^2 2s^1$ | $1s^2 2s^2$ | | | | | | | | | |
| Na | Mg | | | | | | | | | |
| $2p^6 3s^1$ | $2p^6 3s^2$ | | | | | | | | | |
| K | Ca | Sc | Ti | V | Cr | Mn | Fe | Co | Ni | Cu | Zn |
| $3s^2 3p^6 4s^1$ | $3p^6 4s^2$ | $3s^2 3p^6 3d^1 4s^2$ | $3p^6 3d^2 4s^2$ | $3p^6 3d^3 4s^2$ | $3p^6 3d^5 4s^1$ | $3p^6 3d^5 4s^2$ | $3p^6 3d^6 4s^2$ | $3p^6 3d^7 4s^2$ | $3p^6 3d^8 4s^2$ | $3p^6 3d^{10} 4s^1$ | $3p^6 3d^{10} 4s^2$ |
| Rb | Sr | Y | Zr | Nb | Mo | Tc | Ru | Rh | Pd | Ag | Cd |
| $4s^2 4p^6 5s^1$ | $4s^2 4p^6 5s^2$ | $4s^2 4p^6 4d^1 5s^2$ | $4s^2 4p^6 4d^2 5s^2$ | $4s^2 4p^6 4d^3 5s^2$ | $4p^6 4d^5 5s^1$ | $4p^6 4d^5 5s^2$ | $4p^6 4d^6 5s^2$ | $4p^6 4d^7 5s^2$ | $4p^6 4d^8 5s^2$ | $4p^6 4d^{10} 5s^1$ | $4d^{10} 5s^2$ |
| Cs | Ba | La | Hf | Ta | W | Re | Os | Ir | Pt | Au | Hg |
| $5s^2 5p^6 6s^1$ | $5s^2 5p^6 6s^2$ | $5s^2 5p^6 5d^1 6s^2$ | $5p^6 5d^2 6s^2$ | $5p^6 5d^3 6s^2$ | $5p^6 5d^5 6s^1$ | $5p^6 5d^5 6s^2$ | $5p^6 5d^6 6s^2$ | $5d^7 6s^2$ | $5p^6 5d^8 6s^2$ | $5d^{10} 6s^1$ | $5d^{10} 6s^2$ |

**Electronic structure Analyses of solid compounds.** The electronic structures of metal superhydrides are calculated and analyzed by use of several methods, including the Bader's Quantum Theory of Atoms in Molecules (QTAIM) [48], the Electron Localization Function (ELF) (49), the Crystalline Orbital Hamiltonian Population (COHP) and integrated COHP (ICOHP) [50], etc. A systematic study of the Electron Localization Functions at the high symmetry points of metals and their subsets as functions of lattice lengths are shown in Fig. S3 to S10.

**Integrated Charge Difference.** For a given metal, two electron charge densities are calculated, including a self-consistent charge density ($M_{scf}$) and a superposition of atomic charge density ($M_{atom}$). The charge difference is then calculated as $\Delta\rho = \rho(M_{scf}) - \rho(M_{atom})$. The $\Delta\rho=0$ surface divides the crystal space into different regions surrounding the atomic sites and the interstitials. $\Delta\rho$ has positive maxima or negative minima in these regions. The integrated Charge Difference (ICD) are defined for each region by integrating $\Delta\rho$ inside the region. The structures under study may contain one or more types of interstitial quasi-atoms. Their ICDs are calculated separately.

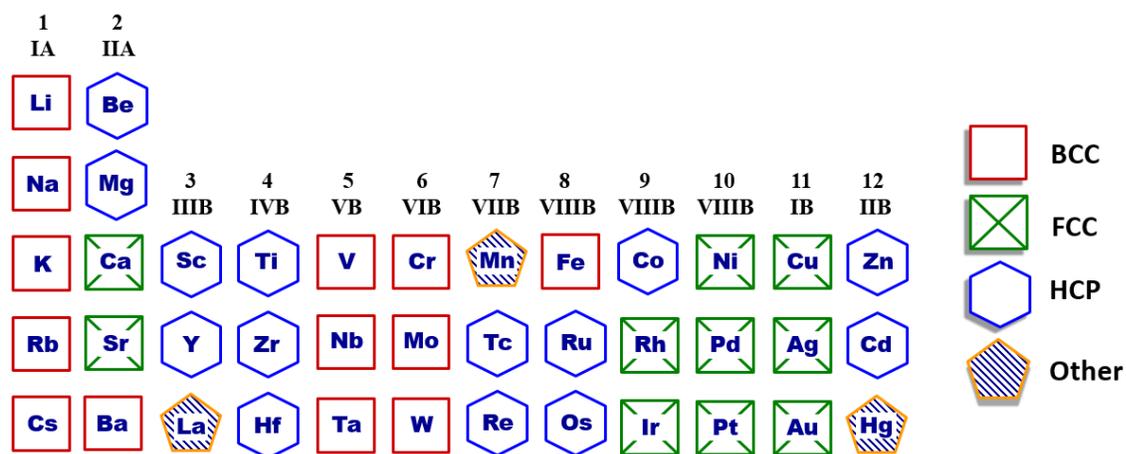

**Fig. S1. Crystal structures of metals at ambient pressure.**

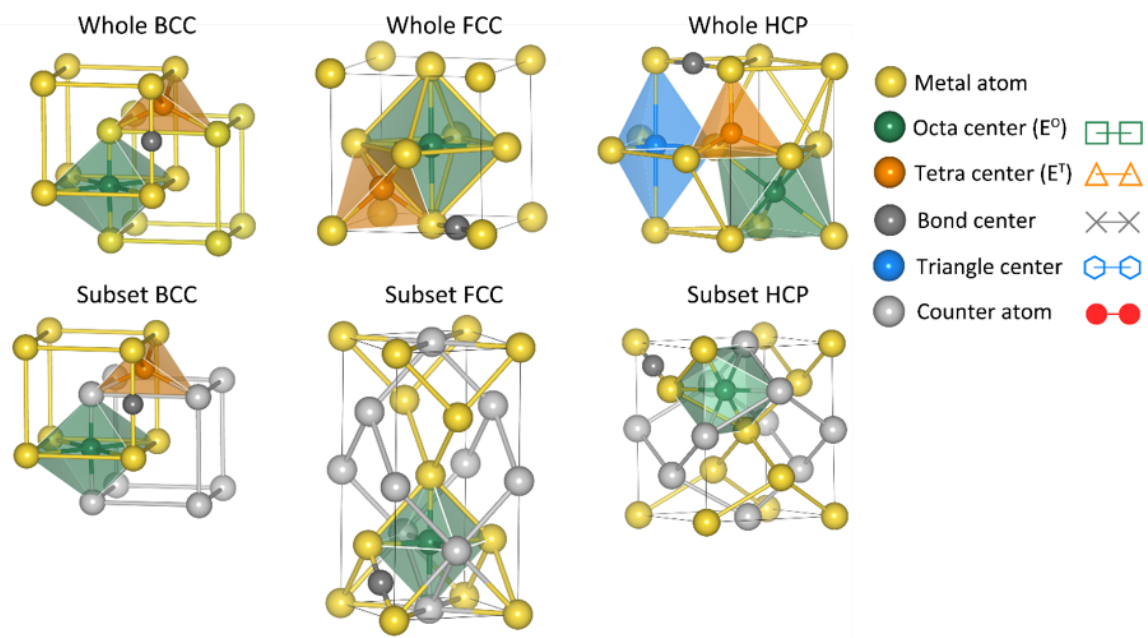

**Fig. S2. Critical positions in whole/subset BCC, whole/subset HCP, and whole/subset FCC lattice.** The symbols on the left denote the critical position in the structures and the right ones represent the corresponding symbols shown in the following ELF graphs.

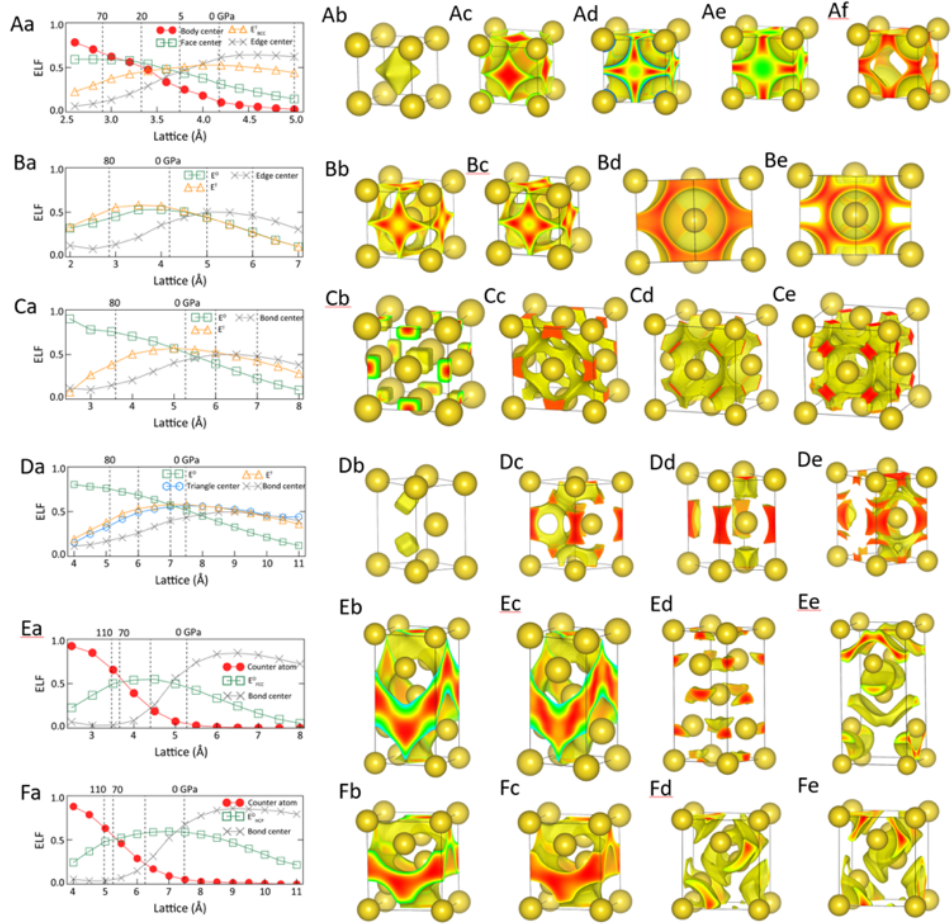

**Fig. S3. Evolution of electron localization in various structures upon the change of volume.**
(**A**) Evolution of electron localization in simple cubic (SC) Na. (**Aa**) ELF values at centers as functions of unit lengths. (**Ab – Af**) ELF of SC Na with the unit lengths of body centered cubic (BCC) Na at 70, 20, 5, 0 and -1.5 (unit length of 5.0 Å) GPa, corresponding to the vertical lines in Aa. B, Evolution of electron localization in BCC Na. (**Ba**) ELF values at centers as functions of unit lengths. (**Bb – Be**) ELF of BCC Na at 80, 0, -1.5 (unit length of 5.0 Å) and -1.1 (unit length of 6.0 Å) GPa, corresponding to the vertical lines in (**Ba**). (**C**) Evolution of electron localization in face centered cubic (FCC) Na. (**Ca**) ELF values at centers as functions of unit lengths. (**Cb – Ce**) ELF of FCC Na at 80, 0, -1.5 (unit length of 6.0 Å) and –1.4 (unit length of 7.0 Å) GPa, corresponding to the vertical lines in (**Ca**). (**D**) Evolution of electron localization in hexagonal close packed (HCP) Na. (**Da**) ELF values at centers as functions of unit lengths. (**Db – De**) ELF of HCP Na at 80, 19 (unit length of 6.0 Å), 2.6 (unit length of 7.0 Å) and 0 GPa, corresponding to the vertical lines in (**Da**). (**E**) Evolution of electron localization in FCC Na subset. (**Ea**) ELF values at centers as functions of unit lengths. (**Eb – Ee**) ELF of FCC Na subset with the unit lengths of FCC Na at 110, 70, 12 (unit length of 4.4 Å) and 0 GPa, corresponding to the vertical lines in (**Ea**). (**F**) Evolution of electron localization in HCP Na subset. (**Fa**) ELF values at centers as functions of unit lengths. (**Fb – Fe**) ELF of HCP Na subset with the unit lengths of HCP Na at 110, 70, 12 (unit length of 6.25 Å) and 0 GPa, corresponding to the vertical lines in (**Fa**).

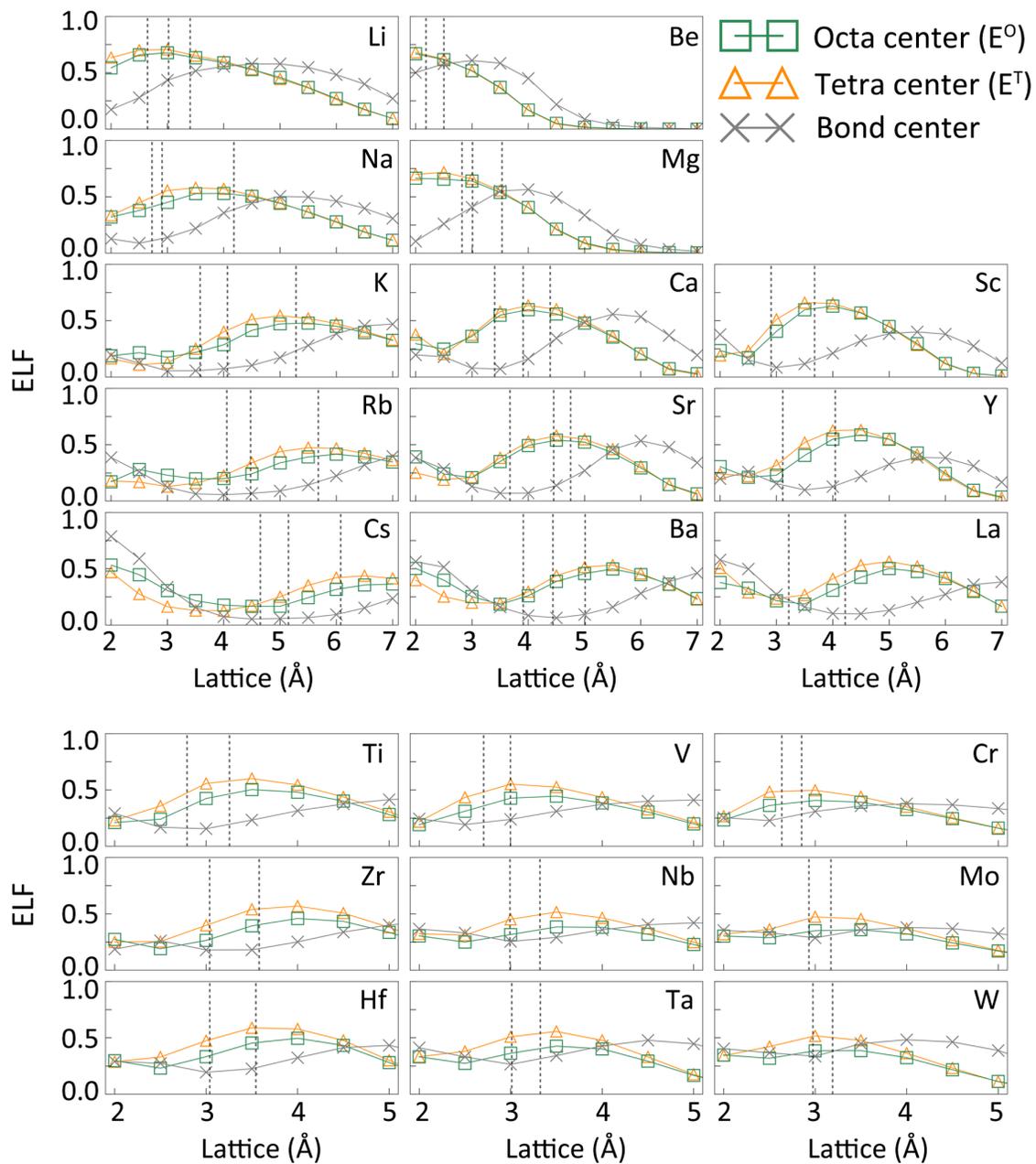

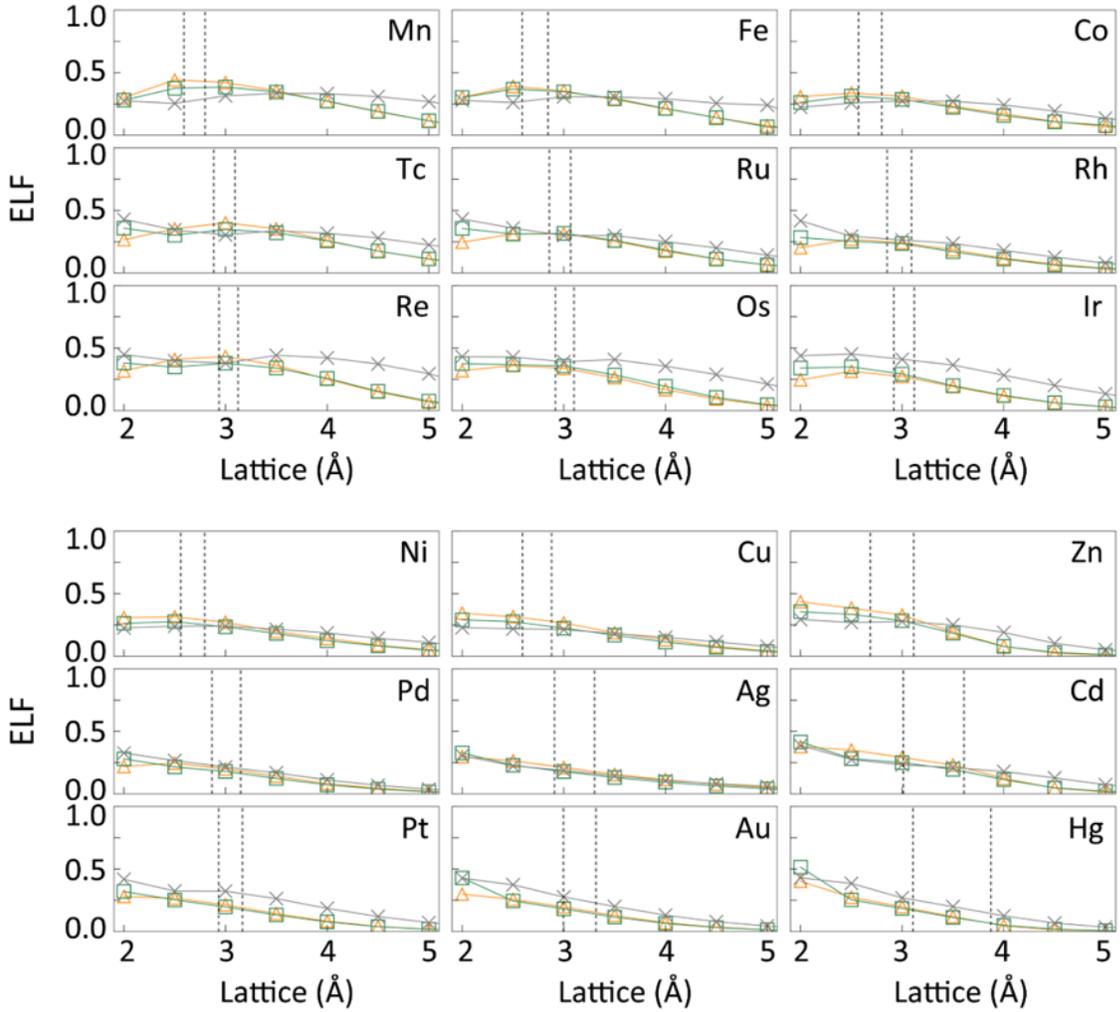

**FIG. S4. ELF values of critical positions in the BCC structure of metals.** In each subgraph, the vertical dash lines from right to left side denotes the decreasing unit lengths with increasing pressure. The applied pressures for Li are 0, 10, 40 GPa; for Na are 0, 70, 120 GPa; for K are 0, 12, 30 GPa; for Rb are 0, 7, 15 GPa; for Cs are 0, 2.5, 5 GPa; for Mg are 0, 50, 100 GPa; for Ca are 0, 10, 40 GPa; for Sr are 0, 3.5, 30 GPa; for Ba are 0, 5.5, 20 GPa; and for other metals are 0, 100 GPa, respectively.

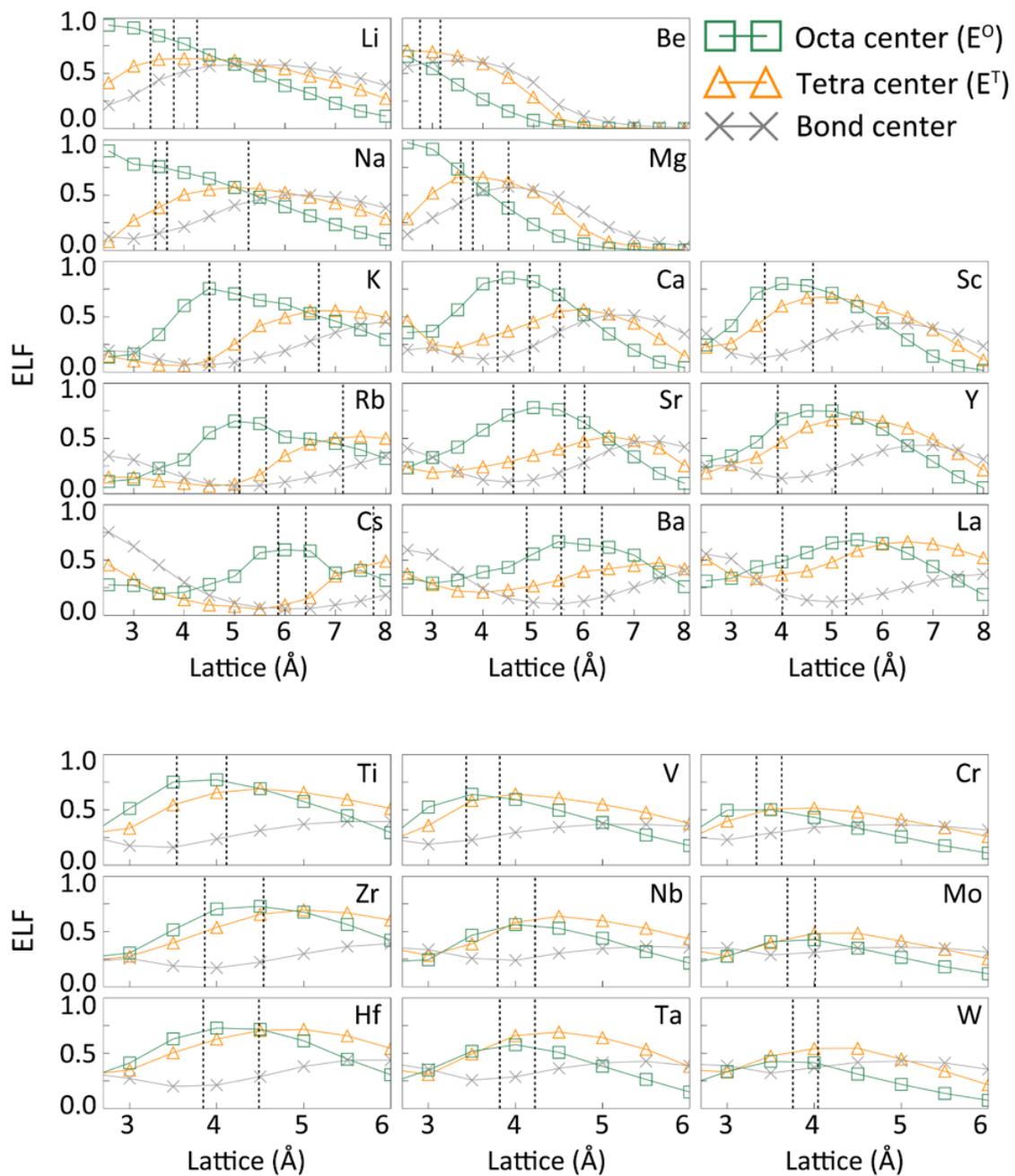

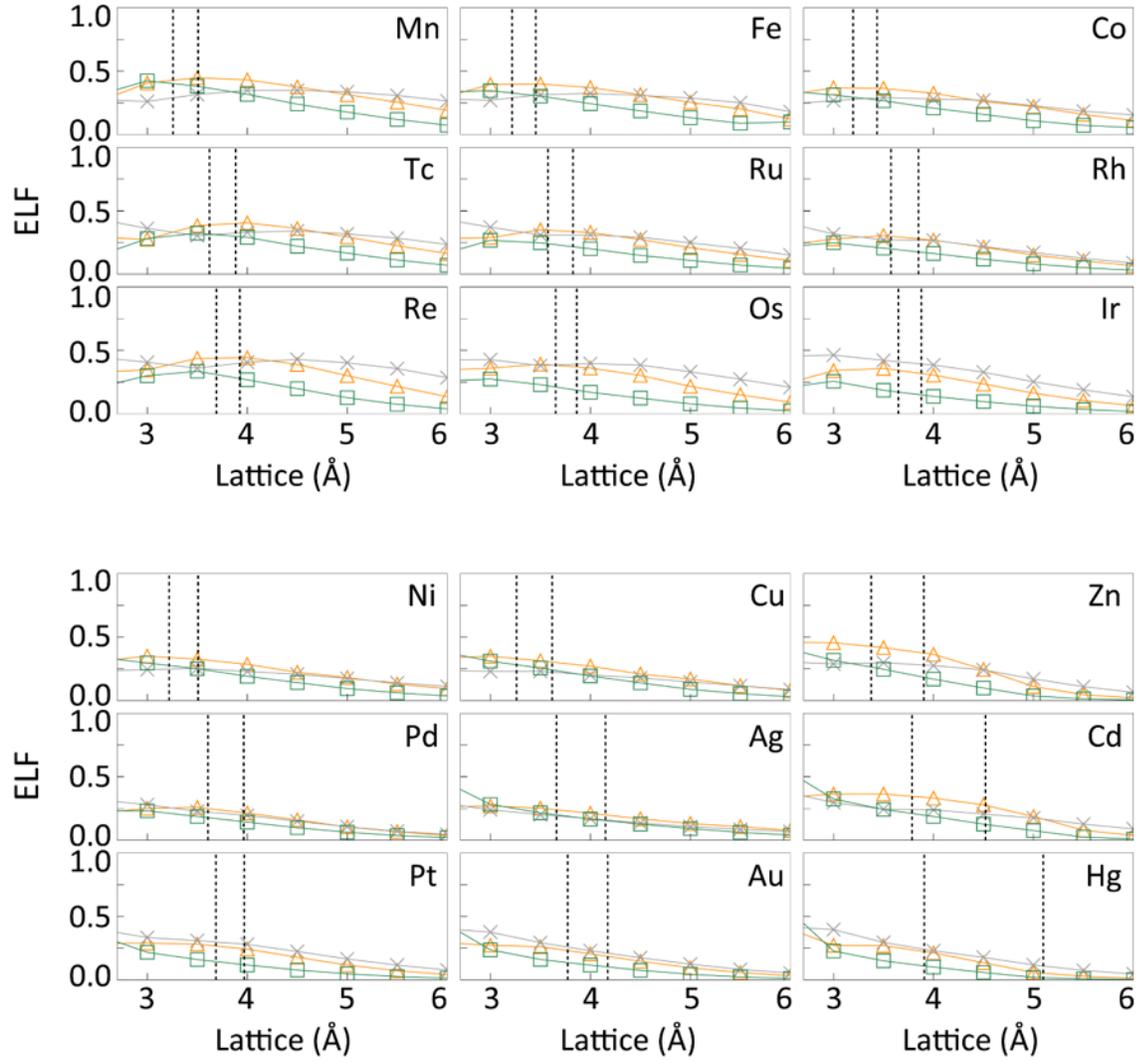

**FIG. S5. ELF values of critical positions in the FCC structure of metals.** In each subgraph, the vertical dash lines from right to left side denotes the decreasing unit lengths with increasing pressure. The applied pressures for Li are 0, 10, 40 GPa; for Na are 0, 70, 120 GPa; for K are 0, 12, 30 GPa; for Rb are 0, 7, 15 GPa; for Cs are 0, 2.5, 5 GPa; for Mg are 0, 50, 100 GPa; for Ca are 0, 10, 40 GPa; for Sr are 0, 3.5, 30 GPa; for Ba are 0, 5.5, 20 GPa; and for other metals are 0, 100 GPa, respectively.

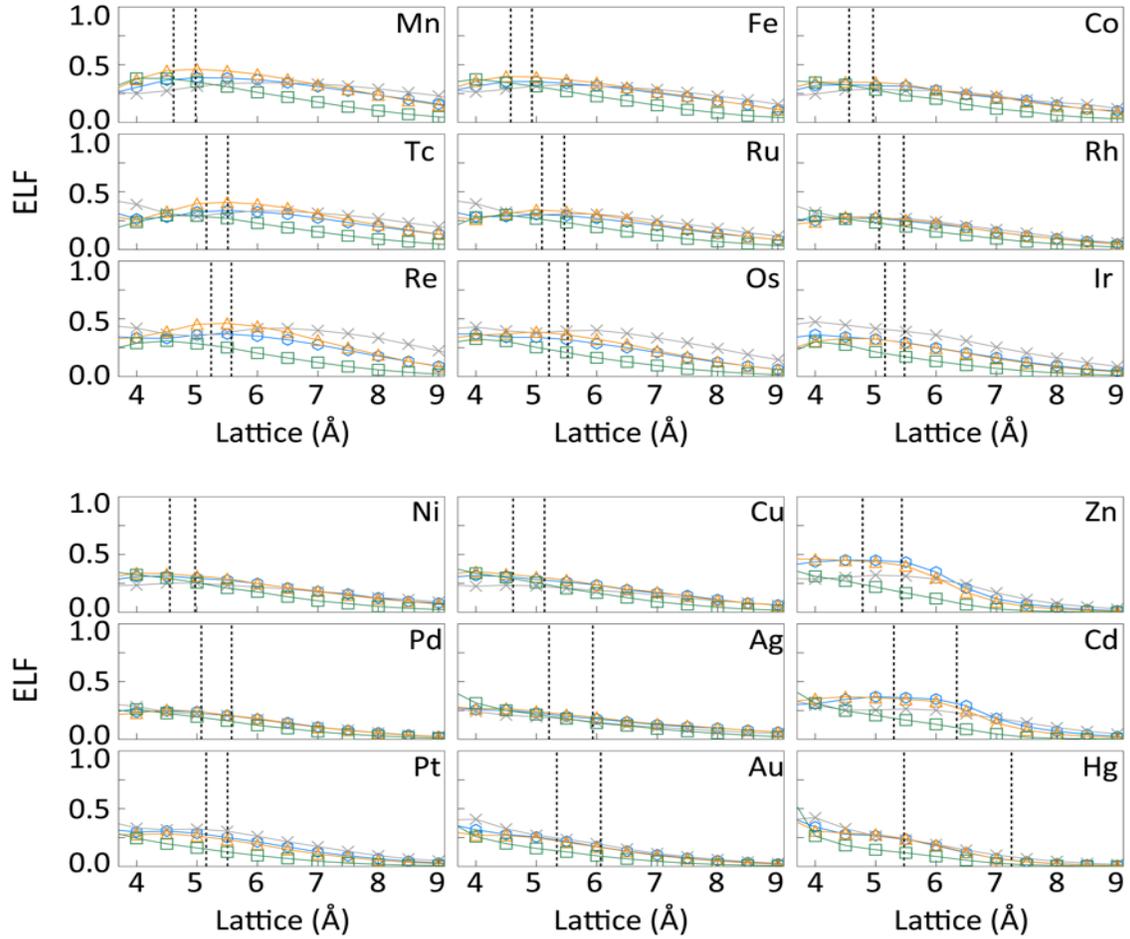

**FIG. S6. ELF values of critical positions in the HCP structure of metals.** In each subgraph, the vertical dash lines from right to left side denotes the decreasing unit lengths with increasing pressure. The applied pressures for Li are 0, 10, 40 GPa; for Na are 0, 70, 120 GPa; for K are 0, 12, 30 GPa; for Rb are 0, 7, 15 GPa; for Cs are 0, 2.5, 5 GPa; for Mg are 0, 50, 100 GPa; for Ca are 0, 10, 40 GPa; for Sr are 0, 3.5, 30 GPa; for Ba are 0, 5.5, 20 GPa; and for other metals are 0, 100 GPa, respectively.

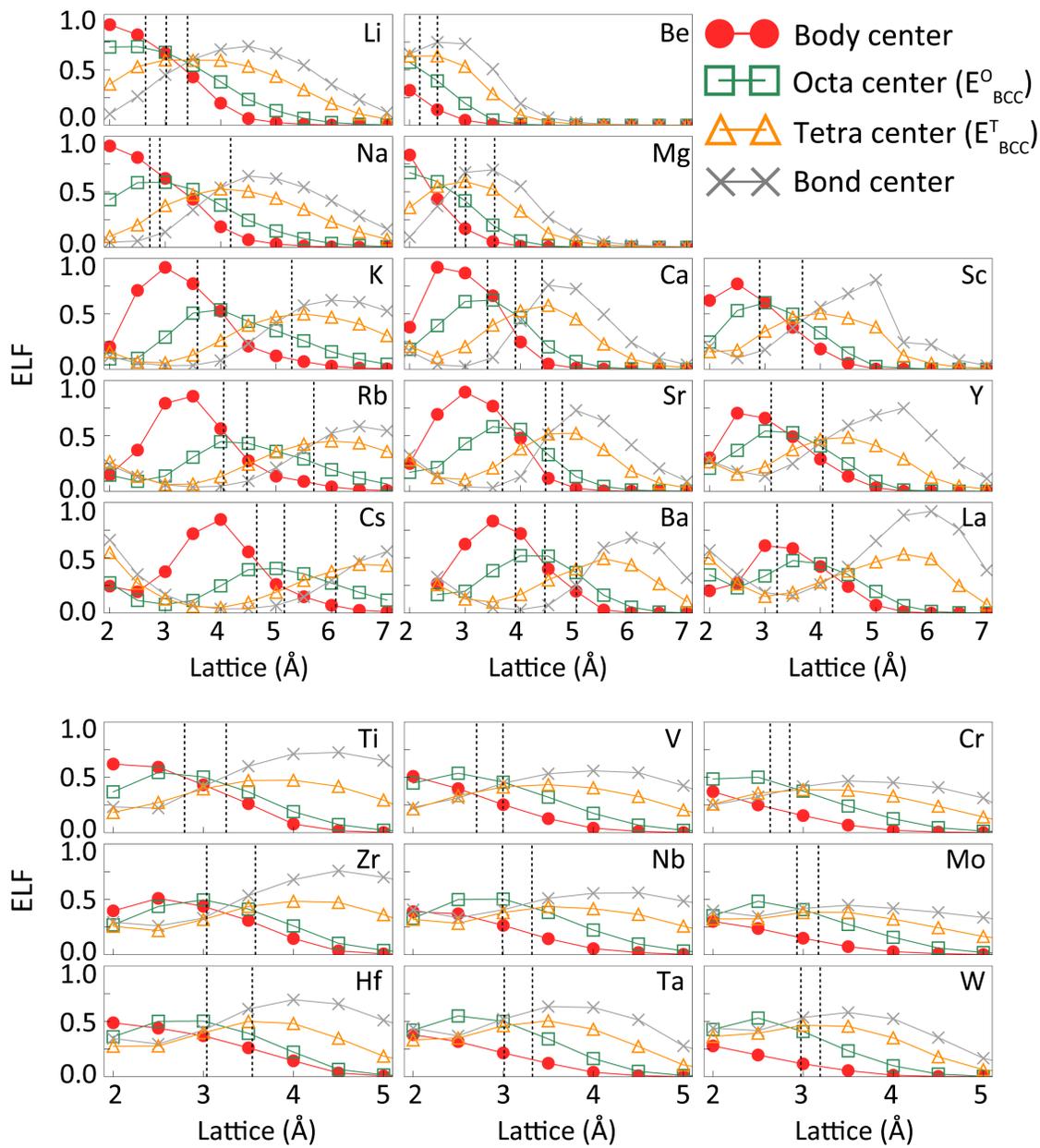

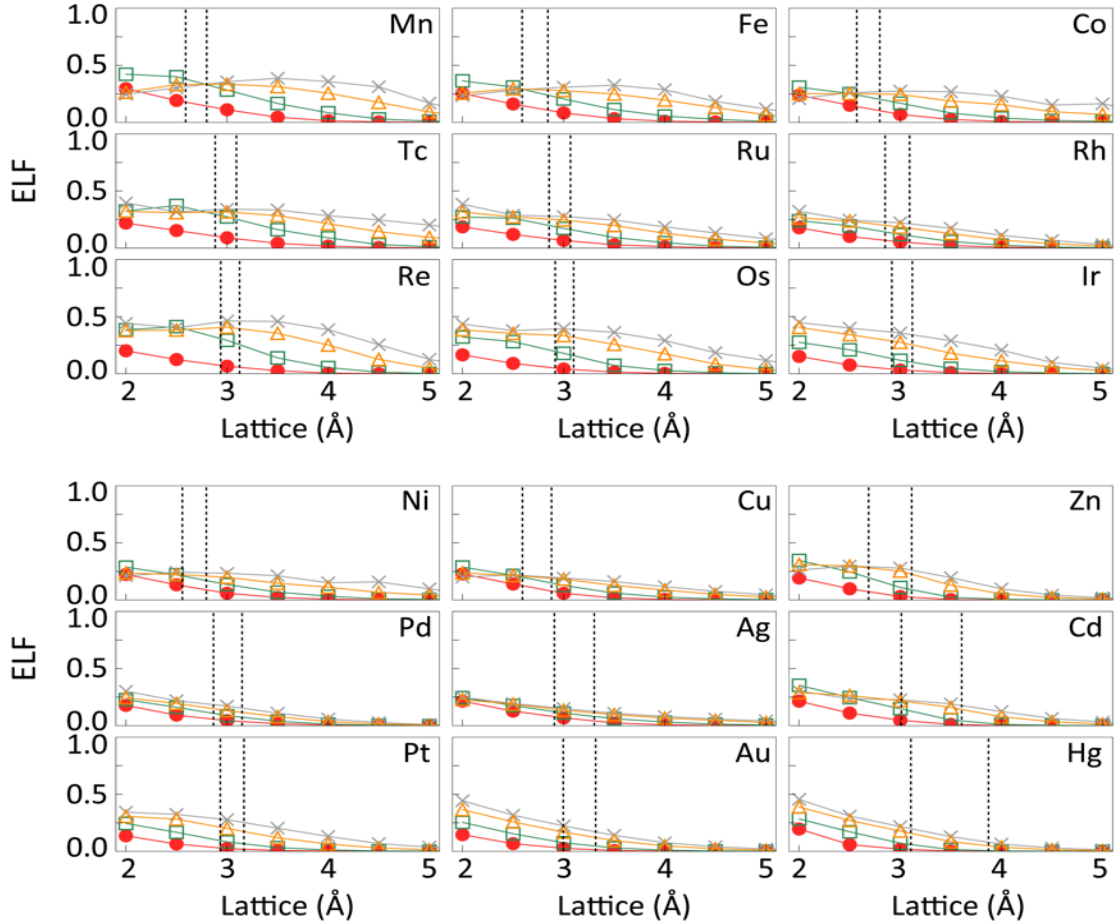

**FIG. S7. ELF values of critical positions in the BCC subset of metals.** In each subgraph, the vertical dash lines from right to left side denotes the decreasing lattice lengths with increasing pressure. The applied pressures for Li are 0, 10, 40 GPa; for Na are 0, 70, 120 GPa; for K are 0, 12, 30 GPa; for Rb are 0, 7, 15 GPa; for Cs are 0, 2.5, 5 GPa; for Mg are 0, 50, 100 GPa; for Ca are 0, 10, 40 GPa; for Sr are 0, 3.5, 30 GPa; for Ba are 0, 5.5, 20 GPa; and for other metals are 0, 100 GPa, respectively.

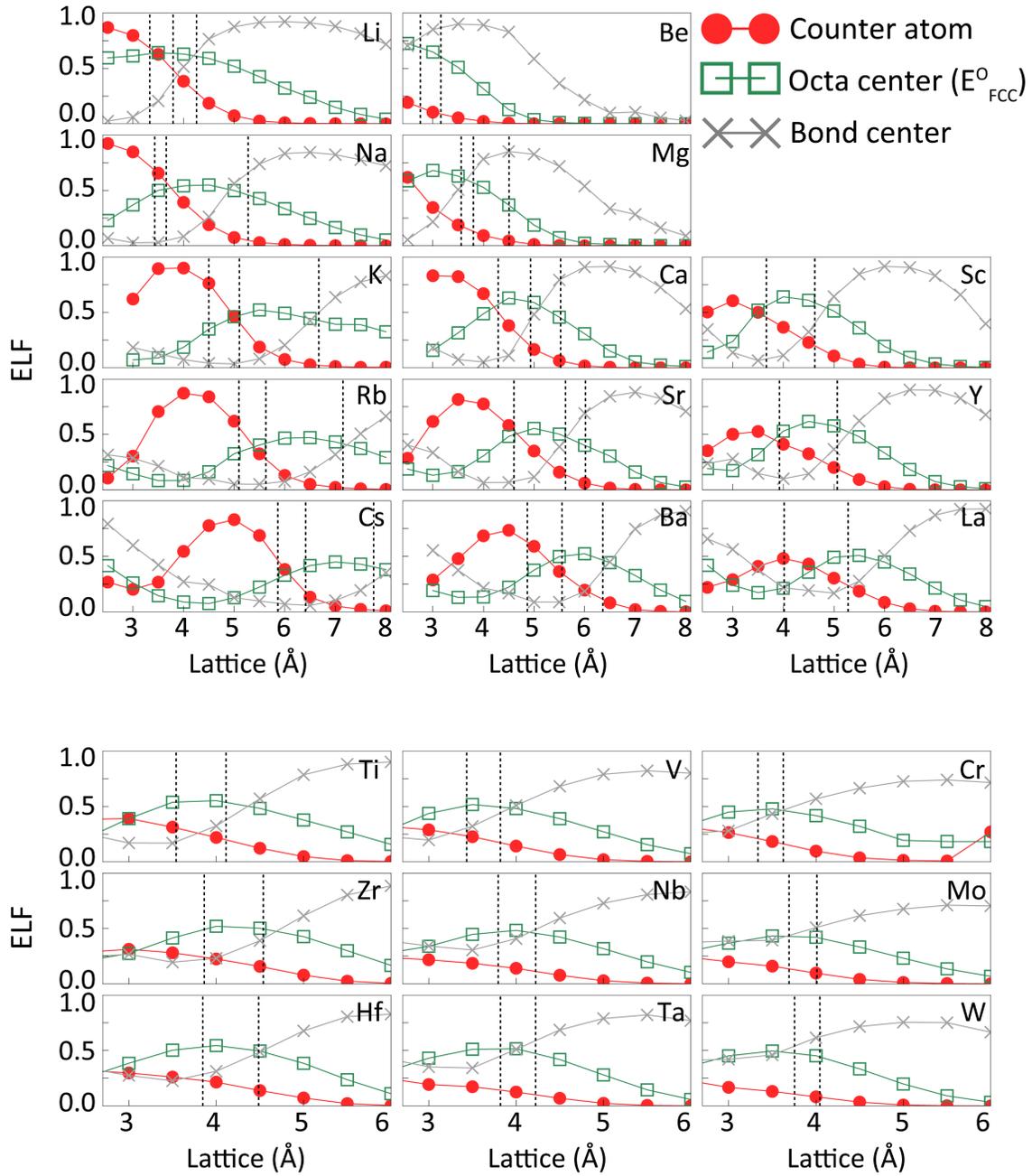

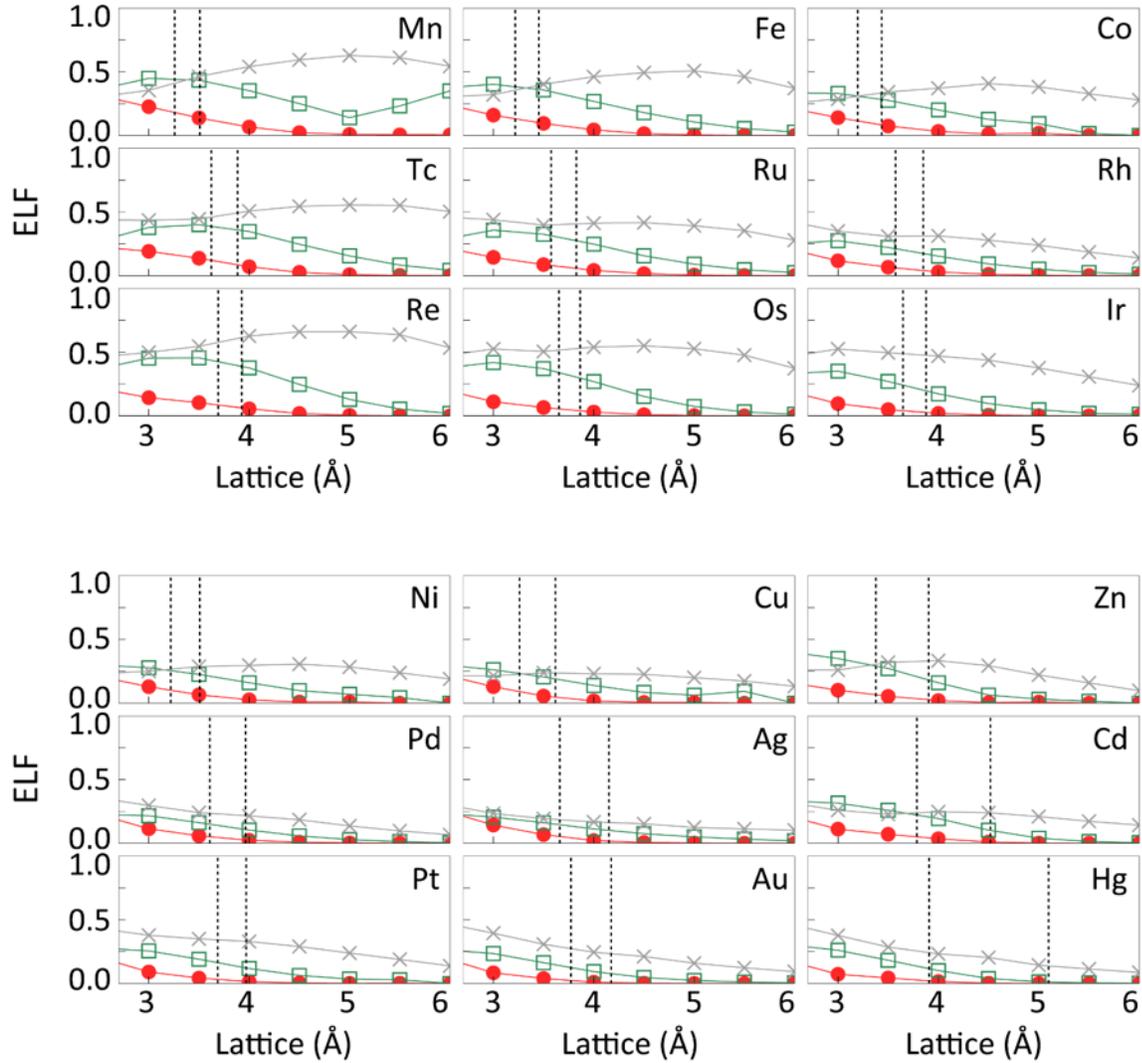

**FIG. S8. ELF values of critical positions in the FCC subset of metals.** In each subgraph, the vertical dash lines from right to left side denotes the decreasing lattice lengths with increasing pressure. The applied pressures for Li are 0, 10, 40 GPa; for Na are 0, 70, 120 GPa; for K are 0, 12, 30 GPa; for Rb are 0, 7, 15 GPa; for Cs are 0, 2.5, 5 GPa; for Mg are 0, 50, 100 GPa; for Ca are 0, 10, 40 GPa; for Sr are 0, 3.5, 30 GPa; for Ba are 0, 5.5, 20 GPa; and for other metals are 0, 100 GPa, respectively.

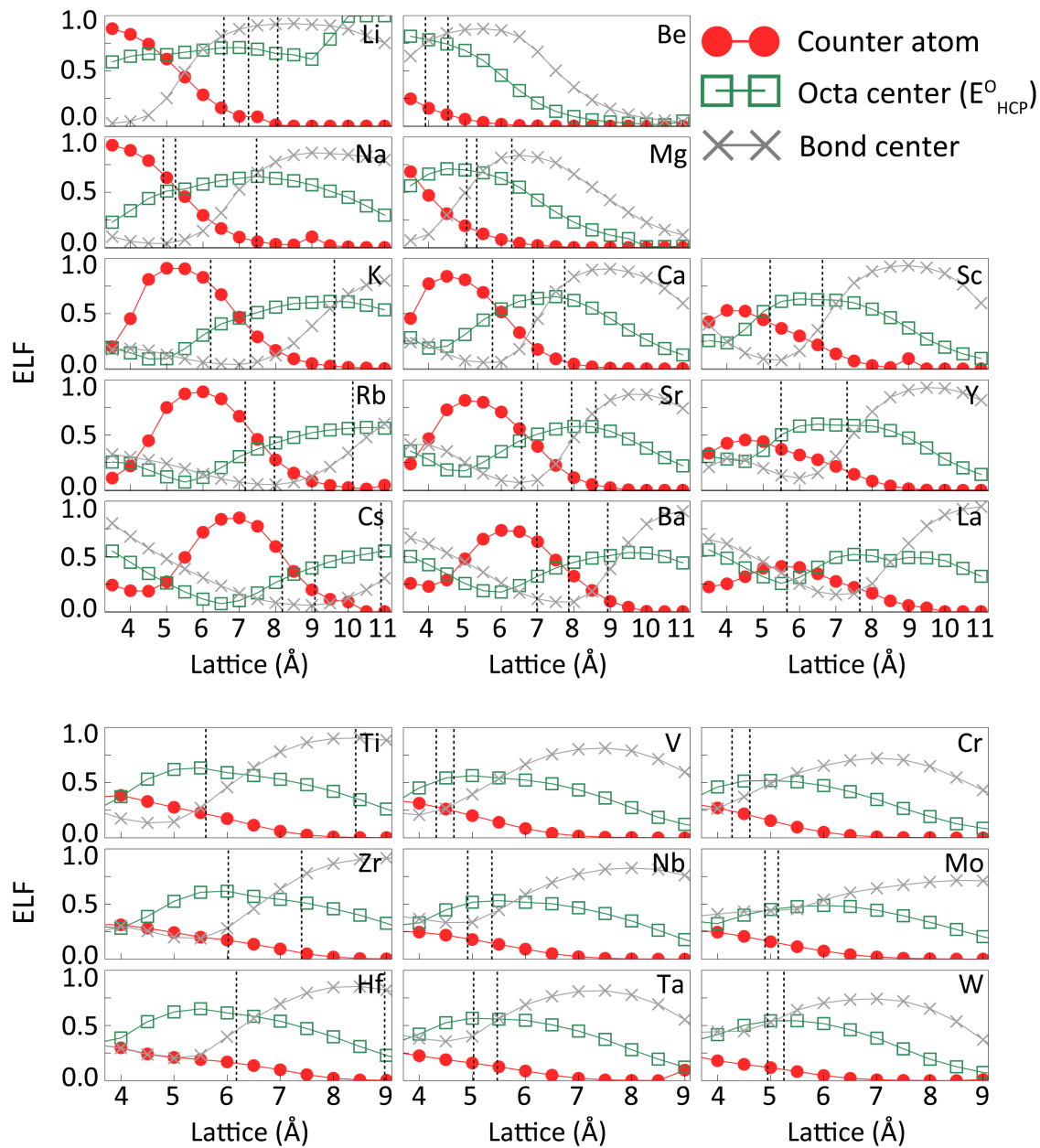

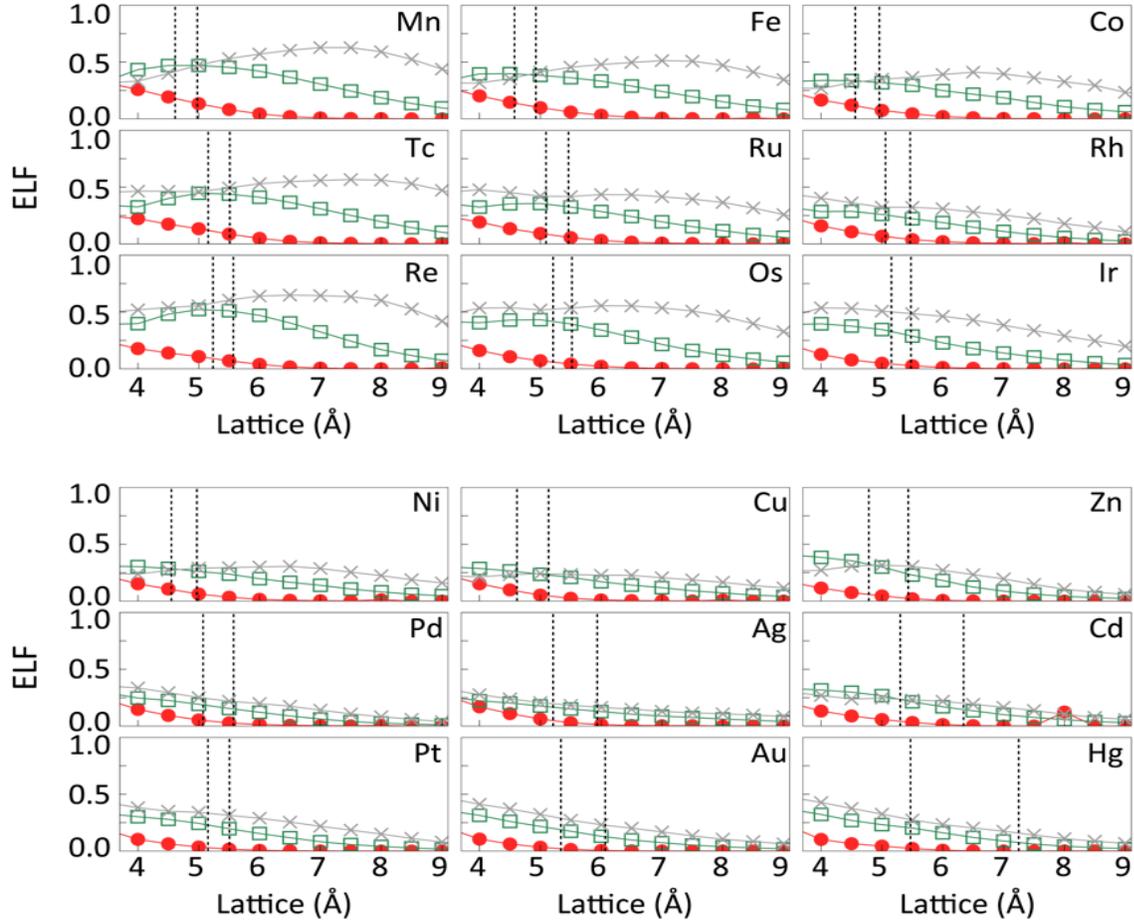

**FIG. S9. ELF values of critical positions in the HCP subset of metals.** In each subgraph, the vertical dash lines from right to left side denotes the decreasing lattice lengths with increasing pressure. The applied pressures for Li are 0, 10, 40 GPa; for Na are 0, 70, 120 GPa; for K are 0, 12, 30 GPa; for Rb are 0, 7, 15 GPa; for Cs are 0, 2.5, 5 GPa; for Mg are 0, 50, 100 GPa; for Ca are 0, 10, 40 GPa; for Sr are 0, 3.5, 30 GPa; for Ba are 0, 5.5, 20 GPa; and for other metals are 0, 100 GPa, respectively.

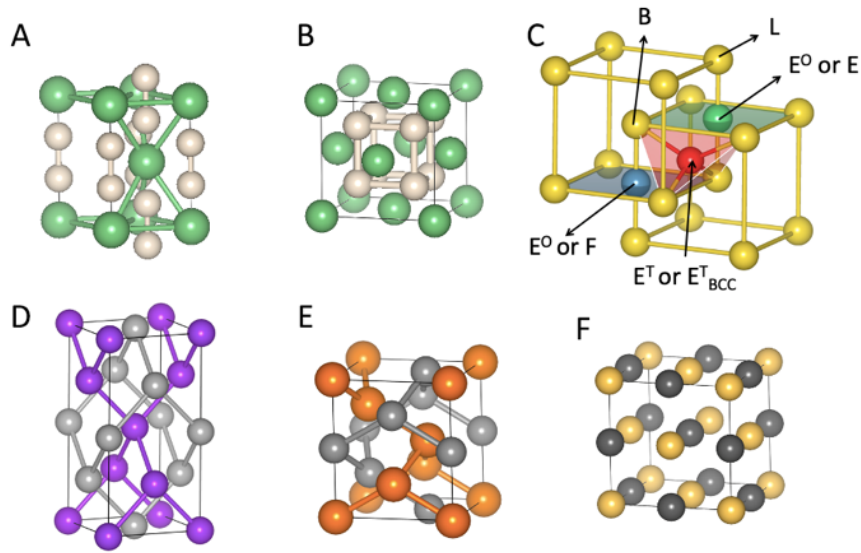

**FIG. S10. Important features of high symmetry structures.** (**A**) The tetrahedral interstitial sites (white balls) form pairs in an HCP lattice. (**B**) The tetrahedral interstitial sites (white balls) form an SC lattice in an FCC lattice. (**C**) The split of a BCC lattice into two SC lattices (subsets). No nearest neighboring atoms in the original BCC lattice are contained in the same SC lattice, showing a perfect partition. The yellow, green, blue, and red balls represent the positions of body centers (B), edge centers (E), face centers (F), and the tetrahedral interstitial sites of the BCC ($E^T_{BCC}$). The lattice sites of the subsets (SC lattices) are also shown by yellow balls. The edge centers (E) and body centers (B) of the subsets are also the octahedral interstitial sites ($E^O$) of the original BCC lattice. (**D**) The split of the FCC lattice into two subsets, represented by purple and grey balls. Among the 12 nearest neighbors in the original FCC lattice, 4 are kept in the same subset as shown by bonds. (**E**) The split of the HCP lattice into two subsets, represented by brown and black balls. Among the 12 nearest neighbors in the original HCP lattice, 4 are kept in the same subset as shown by bonds. (**F**) The split of the SC lattice into two FCC lattices, represented by yellow and black balls. No nearest neighboring atoms in the original SC lattice are contained in the same FCC lattice, showing a perfect partition.

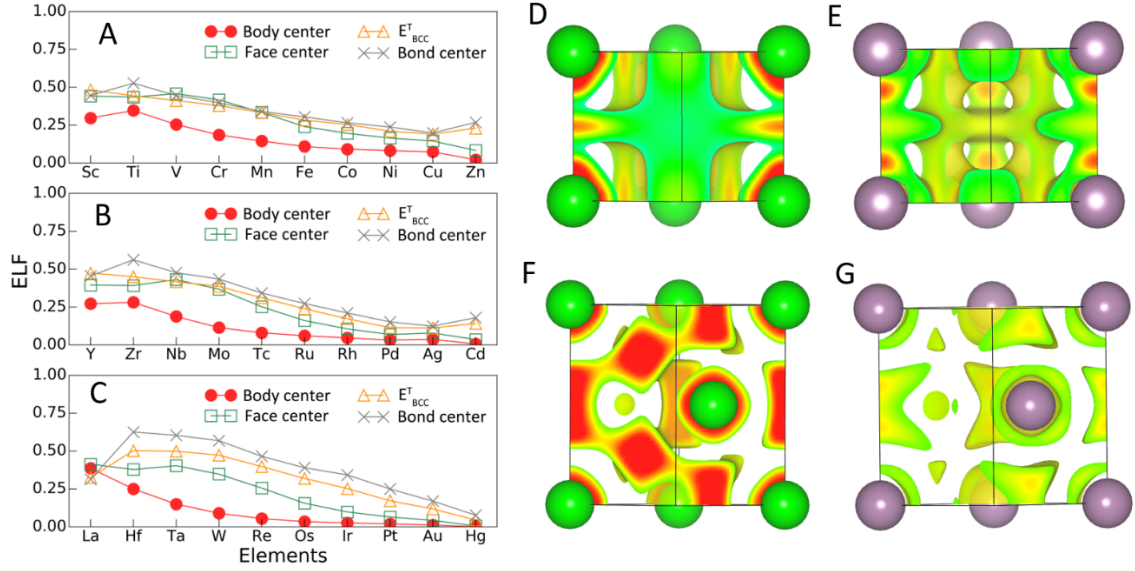

**FIG. S11. Subset interactions in transition metals in BCC structure.** (**A**) ELF values at the high symmetry points in BCC subsets of 3$d$ transition metals. (**B**) ELF values at the high symmetry points in BCC subsets of 4$d$ transition metals. (**C**) ELF values at the high symmetry points in BCC subsets of 5$d$ transition metals. (**D – E**) (110) view of the ELF of Zr and Mo in BCC structure. To compare precisely, the isosurfaces are set at 0.25 and the color code ranges are set as 0 to 0.6, for both metals. (**F – G**) (110) view of the ELF of Zr and Mo in HCP structure. To compare precisely, the isosurfaces are set at 0.35 and the color code ranges are set as 0 to 0.6, for both metals.

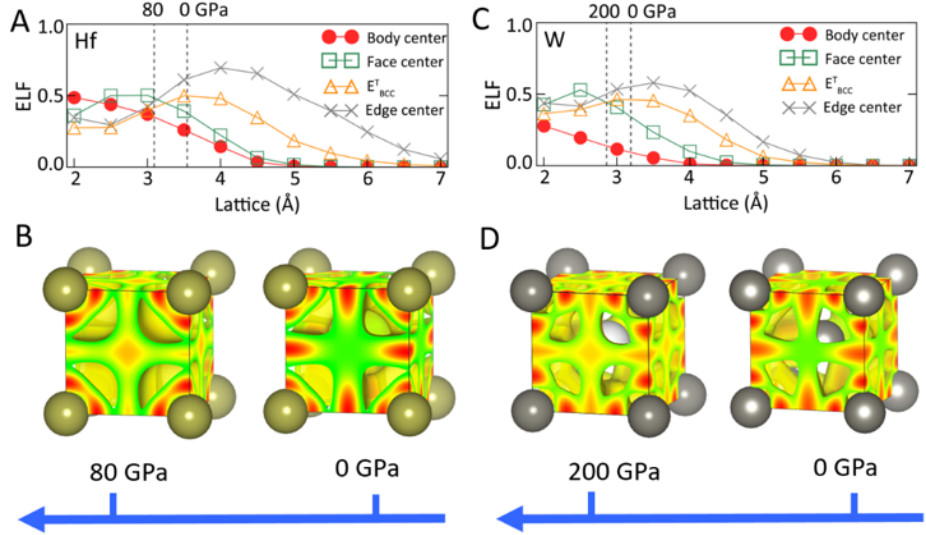

**FIG. S12. Evolution of subset interactions in BCC Hf and W under pressure.** (**A**) ELF of BCC Hf subset (SC) as a function of unit lengths. The vertical lines show the unit lengths of the original BCC Hf at 80 and 0 GPa. (**B**) The schematic of the changes of electron localization in the subsets of BCC Hf under pressure. From right to left, the top row shows the ELF of the BCC subset at 0 and 80 GPa, corresponding to the vertical line in the ELF graph. (**C**) ELF of BCC W subset (SC) as a function of unit lengths. The vertical lines show the unit lengths of the original BCC W at 200 and 0 GPa. (**D**) The schematic of the changes of electron localization in the subsets of BCC W under pressure. From right to left, the top row shows the ELF of the BCC subset at 0 and 200 GPa, corresponding to the vertical line in the ELF graph.

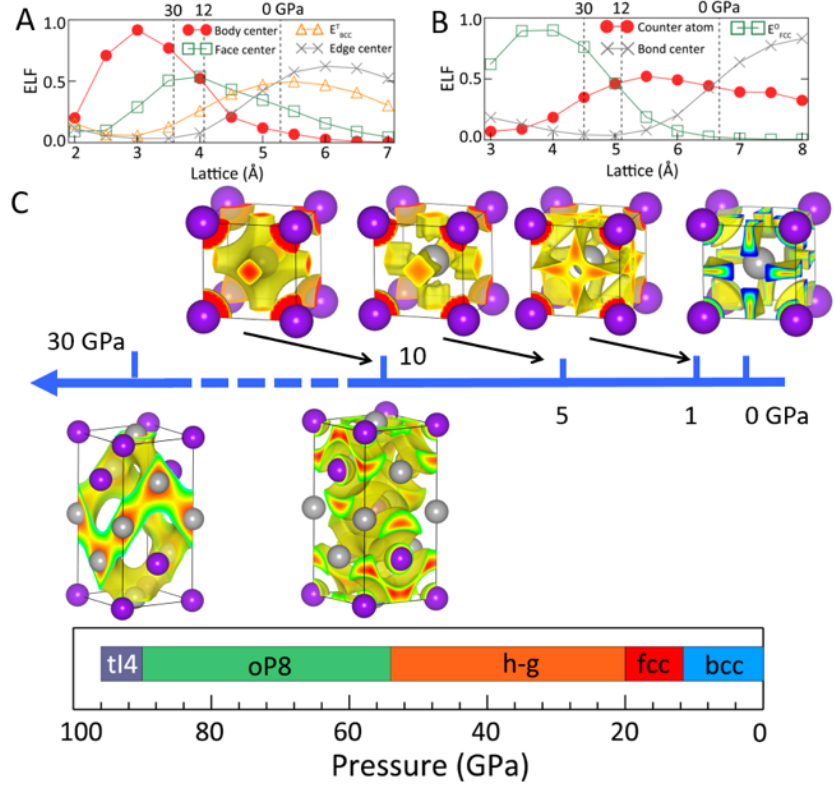

**FIG. S13. Evolution of subset interactions in BCC and FCC K under pressure.** (**A**) ELF of BCC K subset (SC) as a function of unit lengths. The vertical lines show the unit lengths of original BCC K at 30, 12, and 0 GPa. (**B**) ELF of FCC K subset as a function of unit lengths. The vertical lines show the unit lengths of original FCC K at 30, 12, and 0 GPa. (**C**) The schematic of the changes of electron localization in the subsets of BCC and FCC K under pressure. From right to left, the top row shows the ELF of the BCC subset at 10, 5, 1 (perfect matching), and 0 GPa. The bottom row shows the ELF of the FCC subset at 30 and 10 GPa.

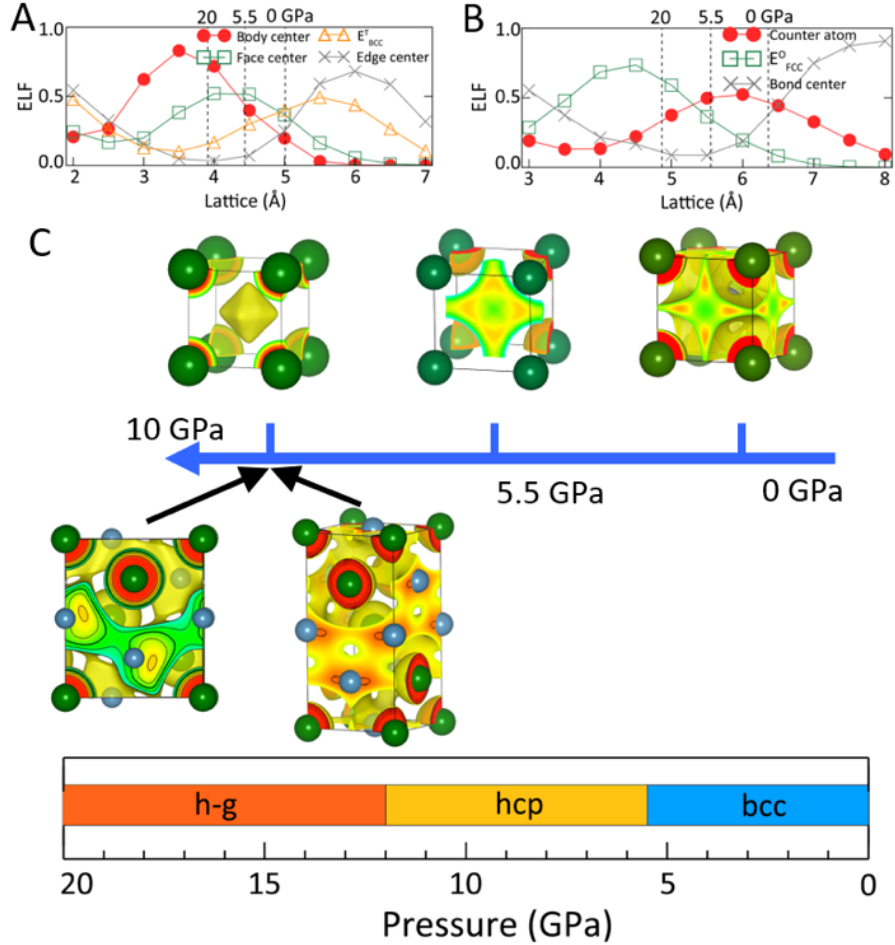

**FIG. S14. Evolution of subset interactions in BCC and FCC Ba under pressure.** (**A**) ELF of BCC Ba subset (SC) as a function of unit lengths. The vertical lines show the unit lengths of the original BCC Ba at 20, 5.5, and 0 GPa. (**B**) ELF of FCC Ba subset as a function of unit lengths. The vertical lines show the unit lengths of the original FCC Ba at 20, 5.5, and 0 GPa. (**C**) The schematic of the changes of electron localization in the subsets of BCC and FCC Ba under pressure. From right to left, the top row shows the ELF of the BCC subset at 0 (perfect matching), 5.5, and 30 GPa. The bottom row shows the ELF of an HCP subset and an FCC subset at 10 GPa. It reveals a large subset repulsion in FCC lattice since the ELF of one subset locates at the atomic positions of another subset. This structure destabilization feature is largely avoided in HCP structure since the ELF maxima of one subset are shifted away from the atomic positions of another subset due to its lower symmetry.

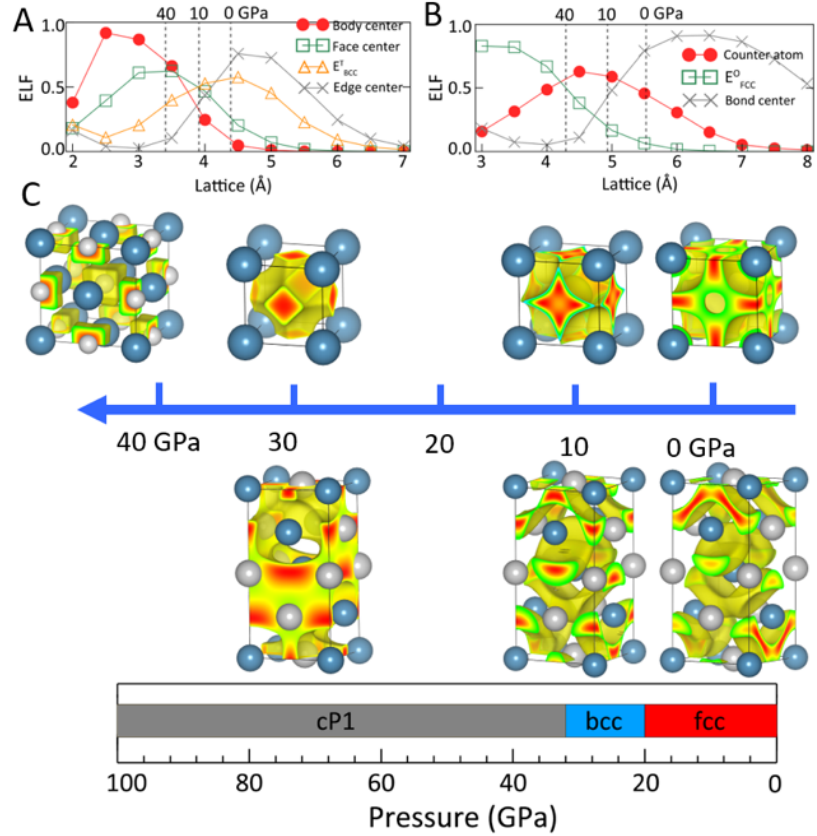

**FIG. S15. Evolution of subset interactions in BCC and FCC Ca under pressure.** (**A**) ELF of BCC Ca subset (SC) as a function of unit lengths. The vertical lines show the unit lengths of original BCC Ca at 30, 10, and 0 GPa. (**B**) ELF of FCC Ca subset as a function of unit lengths. The vertical lines show the unit lengths of original FCC Ca at 30, 10, and 0 GPa. (**C**) The schematic of the changes of electron localization in the subsets of BCC and FCC Ca under pressure. From right to left, the top row shows the ELF of BCC subset at 0, 10 (perfect matching), and 30 GPa, and the ELF of FCC Ca that is the subset of an SC Ca at 40 GPa. The bottom row shows the ELF of the FCC subset at 0, 10, and 30 GPa.

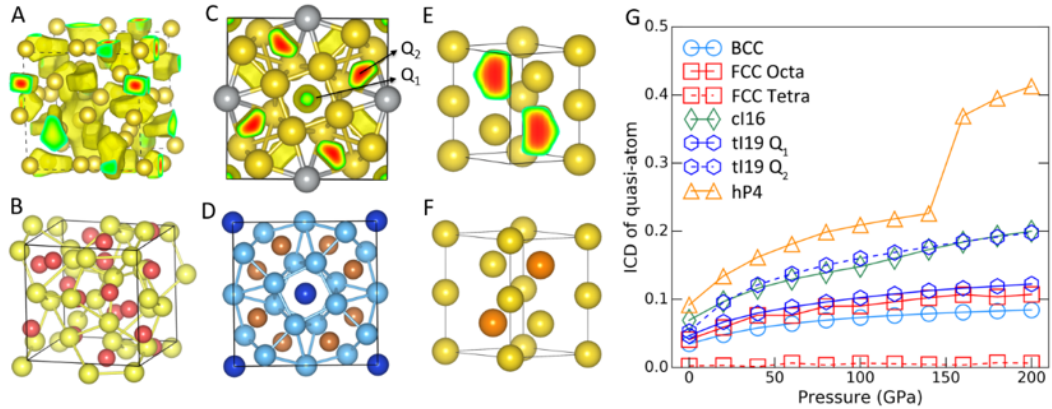

**FIG. S16. The chemistry inside metal structures.** (**A**) The ELF of Na in cI16 structure under 110 GPa. The yellow balls show the positions of the Na atoms. (**B**) The structure of Ba4As3 in $I\bar{4}3d$ structure. The yellow and the red balls represent the Ba and the As atoms. (**C**) The ELF of Na in tI19 structure under 150 GPa. (**D**) Structure of $Ti_5CuSb_2$. The light blue balls represent Ti atoms, the dark blue and the brown balls represent Cu and Sb atoms. (**E**) The ELF of Na in hP4 structure under 200 GPa. (**F**) Structure of $Na_2S$ at 20 GPa. The yellow and brown balls represent the Na and S atoms respectively. (**G**) The calculated Integrated Charge Differences (see methodology section for definition) of quasi-atoms in various structures of Na as functions of pressure.

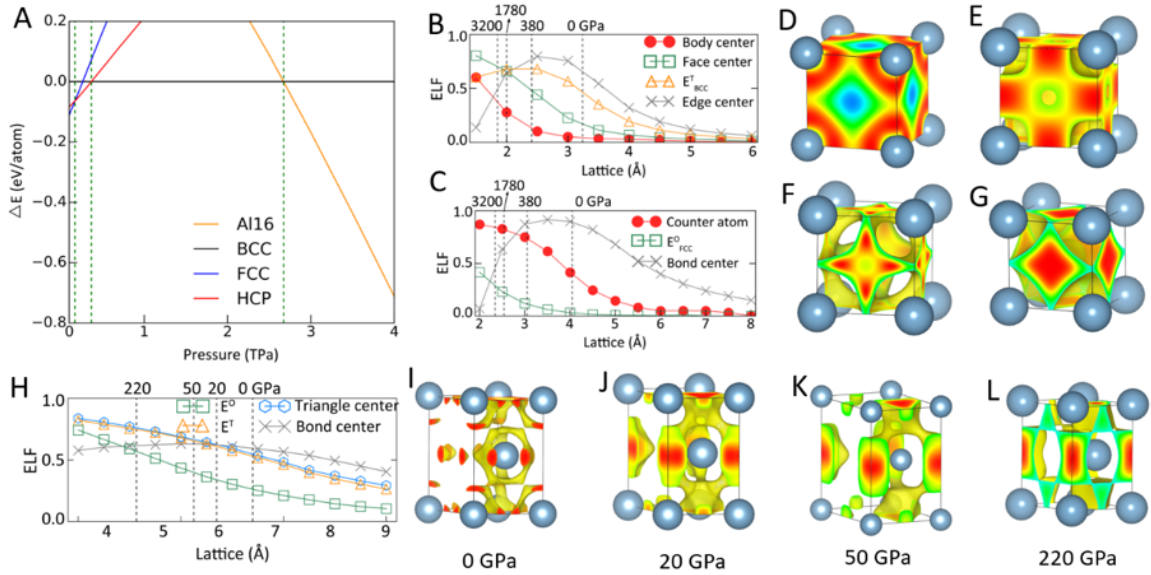

**FIG. S17. Electron localization and structure change of Al under high pressure.** (**A**) Structural evolution of Al under compression up to 4 TPa. (**B**) ELF of BCC Al subset (SC) as a function of unit lengths. The vertical lines show the unit lengths of the original BCC Al at 3200, 1780, 380, and 0 GPa. (**C**) ELF of FCC Al subset (SC) as a function of unit lengths. The vertical lines show the unit lengths of the original FCC Al at 3200, 1780, 380, and 0 GPa. (**D – G**) ELF of BCC Al subset with the unit lengths of BCC Al at 0, 380, 1780, and 3200 GPa, corresponding to the vertical lines in the ELF graph. (**H**) ELF of HCP Al as a function of unit lengths. The vertical lines show the unit lengths of HCP Al at 220, 50, 20, and 0 GPa. (**I – L**) ELF of HCP Al at 0, 20, 50, and 220 GPa, corresponding to the vertical lines in the ELF graph.